\documentstyle[12pt,aaspp4,rotate]{article}
\def\PsfigVersion{1.10}
\def\setDriver{\DvipsDriver} 
\ifx\undefined\psfig\else \fi
\let\LaTeXAtSign=\@
\let\@=\relax
\edef\psfigRestoreAt{\catcode`\@=\number\catcode`@\relax}
\catcode`\@=11\relax
\newwrite\@unused
\def\ps@typeout#1{{\let\protect\string\immediate\write\@unused{#1}}}

\def\DvipsDriver{
	\ps@typeout{psfig/tex \PsfigVersion -dvips}
\def\PsfigSpecials{\DvipsSpecials} 	\def\ps@dir{/}
\def\ps@predir{} }
\def\OzTeXDriver{
	\ps@typeout{psfig/tex \PsfigVersion -oztex}
	\def\PsfigSpecials{\OzTeXSpecials}
	\def\ps@dir{:}
	\def\ps@predir{:}
	\catcode`\^^J=5
}

\def\figurepath{./:}

\def\DoPaths#1{\expandafter\EachPath#1\stoplist}
\def\leer{}
\def\EachPath#1:#2\stoplist{
  \ExistsFile{#1}{\SearchedFile}
  \ifx#2\leer
  \else
    \expandafter\EachPath#2\stoplist
  \fi}
\def\ps@dir{/}
\def\ExistsFile#1#2{%
   \openin1=\ps@predir#1\ps@dir#2
   \ifeof1
       \closein1
   \else
       \closein1
        \ifx\ps@founddir\leer
           \edef\ps@founddir{#1}
        \fi
   \fi}
\def\get@dir#1{%
  \def\ps@founddir{}
  \def\SearchedFile{#1}
  \DoPaths\figurepath
}

\def\@nnil{\@nil}
\def\@empty{}
\def\@psdonoop#1\@@#2#3{}
\def\@psdo#1:=#2\do#3{\edef\@psdotmp{#2}\ifx\@psdotmp\@empty \else
    \expandafter\@psdoloop#2,\@nil,\@nil\@@#1{#3}\fi}
\def\@psdoloop#1,#2,#3\@@#4#5{\def#4{#1}\ifx #4\@nnil \else
       #5\def#4{#2}\ifx #4\@nnil \else#5\@ipsdoloop #3\@@#4{#5}\fi\fi}
\def\@ipsdoloop#1,#2\@@#3#4{\def#3{#1}\ifx #3\@nnil 
       \let\@nextwhile=\@psdonoop \else
      #4\relax\let\@nextwhile=\@ipsdoloop\fi\@nextwhile#2\@@#3{#4}}
\def\@tpsdo#1:=#2\do#3{\xdef\@psdotmp{#2}\ifx\@psdotmp\@empty \else
    \@tpsdoloop#2\@nil\@nil\@@#1{#3}\fi}
\def\@tpsdoloop#1#2\@@#3#4{\def#3{#1}\ifx #3\@nnil 
       \let\@nextwhile=\@psdonoop \else
      #4\relax\let\@nextwhile=\@tpsdoloop\fi\@nextwhile#2\@@#3{#4}}
\ifx\undefined\fbox
\newdimen\fboxrule
\newdimen\fboxsep
\newdimen\ps@tempdima
\newbox\ps@tempboxa
\long\def\fbox#1{\leavevmode\setbox\ps@tempboxa\hbox{#1}\ps@tempdima\fboxrule
    \advance\ps@tempdima \fboxsep \advance\ps@tempdima \dp\ps@tempboxa
   \hbox{\lower \ps@tempdima\hbox
  {\vbox{\hrule height \fboxrule
          \hbox{\vrule width \fboxrule \hskip\fboxsep
          \vbox{\vskip\fboxsep \box\ps@tempboxa\vskip\fboxsep}\hskip 
                 \fboxsep\vrule width \fboxrule}
                 \hrule height \fboxrule}}}}
\fi
\newread\ps@stream
\newif\ifnot@eof       
\newif\if@noisy        
\newif\if@atend        
\newif\if@psfile       
{\catcode`\%=12\global\gdef\epsf@start{
\def\epsf@PS{PS}
\def\epsf@getbb#1{%
\openin\ps@stream=\ps@predir#1
\ifeof\ps@stream\ps@typeout{Error, File #1 not found}\else
   {\not@eoftrue \chardef\other=12
    \def\do##1{\catcode`##1=\other}\dospecials \catcode`\ =10
    \loop
       \if@psfile
	  \read\ps@stream to \epsf@fileline
       \else{
	  \obeyspaces
          \read\ps@stream to \epsf@tmp\global\let\epsf@fileline\epsf@tmp}
       \fi
       \ifeof\ps@stream\not@eoffalse\else
       \if@psfile\else
       \expandafter\epsf@test\epsf@fileline:. \\%
       \fi
          \expandafter\epsf@aux\epsf@fileline:. \\%
       \fi
   \ifnot@eof\repeat
   }\closein\ps@stream\fi}%
\long\def\epsf@test#1#2#3:#4\\{\def\epsf@testit{#1#2}
			\ifx\epsf@testit\epsf@start\else
\ps@typeout{Warning! File does not start with `\epsf@start'.  It may not be a PostScript file.}
			\fi
			\@psfiletrue} 
{\catcode`\%=12\global\let\epsf@percent=
\long\def\epsf@aux#1#2:#3\\{\ifx#1\epsf@percent
   \def\epsf@testit{#2}\ifx\epsf@testit\epsf@bblit
	\@atendfalse
        \epsf@atend #3 . \\%
	\if@atend	
	   \if@verbose{
		\ps@typeout{psfig: found `(atend)'; continuing search}
	   }\fi
        \else
        \epsf@grab #3 . . . \\%
        \not@eoffalse
        \global\no@bbfalse
        \fi
   \fi\fi}%
\def\epsf@grab #1 #2 #3 #4 #5\\{%
   \global\def\epsf@llx{#1}\ifx\epsf@llx\empty
      \epsf@grab #2 #3 #4 #5 .\\\else
   \global\def\epsf@lly{#2}%
   \global\def\epsf@urx{#3}\global\def\epsf@ury{#4}\fi}%
\def\epsf@atendlit{(atend)} 
\def\epsf@atend #1 #2 #3\\{%
   \def\epsf@tmp{#1}\ifx\epsf@tmp\empty
      \epsf@atend #2 #3 .\\\else
   \ifx\epsf@tmp\epsf@atendlit\@atendtrue\fi\fi}

\chardef\psletter = 11 
\chardef\other = 12

\newif \ifdebug 
\newif\ifc@mpute 
\c@mputetrue 

\let\then = \relax
\def\r@dian{pt }
\let\r@dians = \r@dian
\let\dimensionless@nit = \r@dian
\let\dimensionless@nits = \dimensionless@nit
\def\internal@nit{sp }
\let\internal@nits = \internal@nit
\newif\ifstillc@nverging
\def \Mess@ge #1{\ifdebug \then \message {#1} \fi}

{ 
	\catcode `\@ = \psletter
	\gdef \nodimen {\expandafter \n@dimen \the \dimen}
	\gdef \term #1 #2 #3%
	       {\edef \t@ {\the #1}
		\edef \t@@ {\expandafter \n@dimen \the #2\r@dian}%
		\t@rm {\t@} {\t@@} {#3}%
	       }
	\gdef \t@rm #1 #2 #3%
	       {{%
		\count 0 = 0
		\dimen 0 = 1 \dimensionless@nit
		\dimen 2 = #2\relax
		\Mess@ge {Calculating term #1 of \nodimen 2}%
		\loop
		\ifnum	\count 0 < #1
		\then	\advance \count 0 by 1
			\Mess@ge {Iteration \the \count 0 \space}%
			\Multiply \dimen 0 by {\dimen 2}%
			\Mess@ge {After multiplication, term = \nodimen 0}%
			\Divide \dimen 0 by {\count 0}%
			\Mess@ge {After division, term = \nodimen 0}%
		\repeat
		\Mess@ge {Final value for term #1 of 
				\nodimen 2 \space is \nodimen 0}%
		\xdef \Term {#3 = \nodimen 0 \r@dians}%
		\aftergroup \Term
	       }}
	\catcode `\p = \other
	\catcode `\t = \other
	\gdef \n@dimen #1pt{#1} 
}

\def \Divide #1by #2{\divide #1 by #2} 

\def \Multiply #1by #2
       {{
	\count 0 = #1\relax
	\count 2 = #2\relax
	\count 4 = 65536
	\Mess@ge {Before scaling, count 0 = \the \count 0 \space and
			count 2 = \the \count 2}%
	\ifnum	\count 0 > 32767 
	\then	\divide \count 0 by 4
		\divide \count 4 by 4
	\else	\ifnum	\count 0 < -32767
		\then	\divide \count 0 by 4
			\divide \count 4 by 4
		\else
		\fi
	\fi
	\ifnum	\count 2 > 32767 
	\then	\divide \count 2 by 4
		\divide \count 4 by 4
	\else	\ifnum	\count 2 < -32767
		\then	\divide \count 2 by 4
			\divide \count 4 by 4
		\else
		\fi
	\fi
	\multiply \count 0 by \count 2
	\divide \count 0 by \count 4
	\xdef \product {#1 = \the \count 0 \internal@nits}%
	\aftergroup \product
       }}

\def\r@duce{\ifdim\dimen0 > 90\r@dian \then   
		\multiply\dimen0 by -1
		\advance\dimen0 by 180\r@dian
		\r@duce
	    \else \ifdim\dimen0 < -90\r@dian \then  
		\advance\dimen0 by 360\r@dian
		\r@duce
		\fi
	    \fi}

\def\Sine#1%
       {{%
	\dimen 0 = #1 \r@dian
	\r@duce
	\ifdim\dimen0 = -90\r@dian \then
	   \dimen4 = -1\r@dian
	   \c@mputefalse
	\fi
	\ifdim\dimen0 = 90\r@dian \then
	   \dimen4 = 1\r@dian
	   \c@mputefalse
	\fi
	\ifdim\dimen0 = 0\r@dian \then
	   \dimen4 = 0\r@dian
	   \c@mputefalse
	\fi
	\ifc@mpute \then
		\divide\dimen0 by 180
		\dimen0=3.141592654\dimen0
		\dimen 2 = 3.1415926535897963\r@dian 
		\divide\dimen 2 by 2 
		\Mess@ge {Sin: calculating Sin of \nodimen 0}%
		\count 0 = 1 
		\dimen 2 = 1 \r@dian 
		\dimen 4 = 0 \r@dian 
		\loop
			\ifnum	\dimen 2 = 0 
			\then	\stillc@nvergingfalse 
			\else	\stillc@nvergingtrue
			\fi
			\ifstillc@nverging 
			\then	\term {\count 0} {\dimen 0} {\dimen 2}%
				\advance \count 0 by 2
				\count 2 = \count 0
				\divide \count 2 by 2
				\ifodd	\count 2 
				\then	\advance \dimen 4 by \dimen 2
				\else	\advance \dimen 4 by -\dimen 2
				\fi
		\repeat
	\fi		
			\xdef \sine {\nodimen 4}%
       }}

\def\Cosine#1{\ifx\sine\UnDefined\edef\Savesine{\relax}\else
		             \edef\Savesine{\sine}\fi
	{\dimen0=#1\r@dian\advance\dimen0 by 90\r@dian
	 \Sine{\nodimen 0}
	 \xdef\cosine{\sine}
	 \xdef\sine{\Savesine}}}	      

\def\psdraft{
	\def\@psdraft{0}
}
\def\psfull{
	\def\@psdraft{100}
}

\psfull

\newif\if@scalefirst
\def\psscalefirst{\@scalefirsttrue}
\def\psrotatefirst{\@scalefirstfalse}
\psrotatefirst

\newif\if@draftbox
\def\psnodraftbox{
	\@draftboxfalse
}
\def\psdraftbox{
	\@draftboxtrue
}
\@draftboxtrue

\newif\if@prologfile
\newif\if@postlogfile
\def\pssilent{
	\@noisyfalse
}
\def\psnoisy{
	\@noisytrue
}
\psnoisy
\newif\if@bbllx
\newif\if@bblly
\newif\if@bburx
\newif\if@bbury
\newif\if@height
\newif\if@width
\newif\if@rheight
\newif\if@rwidth
\newif\if@angle
\newif\if@clip
\newif\if@verbose
\def\@p@@sclip#1{\@cliptrue}
\newif\if@decmpr
\def\@p@@sfigure#1{\def\@p@sfile{null}\def\@p@sbbfile{null}\@decmprfalse
   \openin1=\ps@predir#1
   \ifeof1
	\closein1
	\get@dir{#1}
	\ifx\ps@founddir\leer
		\openin1=\ps@predir#1.bb
		\ifeof1
			\closein1
			\get@dir{#1.bb}
			\ifx\ps@founddir\leer
				\ps@typeout{Can't find #1 in \figurepath}
			\else
				\@decmprtrue
				\def\@p@sfile{\ps@founddir\ps@dir#1}
				\def\@p@sbbfile{\ps@founddir\ps@dir#1.bb}
			\fi
		\else
			\closein1
			\@decmprtrue
			\def\@p@sfile{#1}
			\def\@p@sbbfile{#1.bb}
		\fi
	\else
		\def\@p@sfile{\ps@founddir\ps@dir#1}
		\def\@p@sbbfile{\ps@founddir\ps@dir#1}
	\fi
   \else
	\closein1
	\def\@p@sfile{#1}
	\def\@p@sbbfile{#1}
   \fi
}
\def\@p@@sfile#1{\@p@@sfigure{#1}}
\def\@p@@sbbllx#1{
		\@bbllxtrue
		\dimen100=#1
		\edef\@p@sbbllx{\number\dimen100}
}
\def\@p@@sbblly#1{
		\@bbllytrue
		\dimen100=#1
		\edef\@p@sbblly{\number\dimen100}
}
\def\@p@@sbburx#1{
		\@bburxtrue
		\dimen100=#1
		\edef\@p@sbburx{\number\dimen100}
}
\def\@p@@sbbury#1{
		\@bburytrue
		\dimen100=#1
		\edef\@p@sbbury{\number\dimen100}
}
\def\@p@@sheight#1{
		\@heighttrue
		\dimen100=#1
   		\edef\@p@sheight{\number\dimen100}
}
\def\@p@@swidth#1{
		\@widthtrue
		\dimen100=#1
		\edef\@p@swidth{\number\dimen100}
}
\def\@p@@srheight#1{
		\@rheighttrue
		\dimen100=#1
		\edef\@p@srheight{\number\dimen100}
}
\def\@p@@srwidth#1{
		\@rwidthtrue
		\dimen100=#1
		\edef\@p@srwidth{\number\dimen100}
}
\def\@p@@sangle#1{
		\@angletrue
		\edef\@p@sangle{#1} 
}
\def\@p@@ssilent#1{ 
		\@verbosefalse
}
\def\@p@@sprolog#1{\@prologfiletrue\def\@prologfileval{#1}}
\def\@p@@spostlog#1{\@postlogfiletrue\def\@postlogfileval{#1}}
\def\@cs@name#1{\csname #1\endcsname}
\def\@setparms#1=#2,{\@cs@name{@p@@s#1}{#2}}
\def\ps@init@parms{
		\@bbllxfalse \@bbllyfalse
		\@bburxfalse \@bburyfalse
		\@heightfalse \@widthfalse
		\@rheightfalse \@rwidthfalse
		\def\@p@sbbllx{}\def\@p@sbblly{}
		\def\@p@sbburx{}\def\@p@sbbury{}
		\def\@p@sheight{}\def\@p@swidth{}
		\def\@p@srheight{}\def\@p@srwidth{}
		\def\@p@sangle{0}
		\def\@p@sfile{} \def\@p@sbbfile{}
		\def\@p@scost{10}
		\def\@sc{}
		\@prologfilefalse
		\@postlogfilefalse
		\@clipfalse
		\if@noisy
			\@verbosetrue
		\else
			\@verbosefalse
		\fi
}
\def\parse@ps@parms#1{
	 	\@psdo\@psfiga:=#1\do
		   {\expandafter\@setparms\@psfiga,}}
\newif\ifno@bb
\def\bb@missing{
	\if@verbose{
		\ps@typeout{psfig: searching \@p@sbbfile \space  for bounding box}
	}\fi
	\no@bbtrue
	\epsf@getbb{\@p@sbbfile}
        \ifno@bb \else \bb@cull\epsf@llx\epsf@lly\epsf@urx\epsf@ury\fi
}	
\def\bb@cull#1#2#3#4{
	\dimen100=#1 bp\edef\@p@sbbllx{\number\dimen100}
	\dimen100=#2 bp\edef\@p@sbblly{\number\dimen100}
	\dimen100=#3 bp\edef\@p@sbburx{\number\dimen100}
	\dimen100=#4 bp\edef\@p@sbbury{\number\dimen100}
	\no@bbfalse
}
\newdimen\p@intvaluex
\newdimen\p@intvaluey
\def\rotate@#1#2{{\dimen0=#1 sp\dimen1=#2 sp
		  \global\p@intvaluex=\cosine\dimen0
		  \dimen3=\sine\dimen1
		  \global\advance\p@intvaluex by -\dimen3
		  \global\p@intvaluey=\sine\dimen0
		  \dimen3=\cosine\dimen1
		  \global\advance\p@intvaluey by \dimen3
		  }}
\def\compute@bb{
		\no@bbfalse
		\if@bbllx \else \no@bbtrue \fi
		\if@bblly \else \no@bbtrue \fi
		\if@bburx \else \no@bbtrue \fi
		\if@bbury \else \no@bbtrue \fi
		\ifno@bb \bb@missing \fi
		\ifno@bb \ps@typeout{FATAL ERROR: no bb supplied or found}
			\no-bb-error
		\fi
		\count203=\@p@sbburx
		\count204=\@p@sbbury
		\advance\count203 by -\@p@sbbllx
		\advance\count204 by -\@p@sbblly
		\edef\ps@bbw{\number\count203}
		\edef\ps@bbh{\number\count204}
		\if@angle 
			\Sine{\@p@sangle}\Cosine{\@p@sangle}
	        	{\dimen100=\maxdimen\xdef\r@p@sbbllx{\number\dimen100}
					    \xdef\r@p@sbblly{\number\dimen100}
			                    \xdef\r@p@sbburx{-\number\dimen100}
					    \xdef\r@p@sbbury{-\number\dimen100}}
                        \def\minmaxtest{
			   \ifnum\number\p@intvaluex<\r@p@sbbllx
			      \xdef\r@p@sbbllx{\number\p@intvaluex}\fi
			   \ifnum\number\p@intvaluex>\r@p@sbburx
			      \xdef\r@p@sbburx{\number\p@intvaluex}\fi
			   \ifnum\number\p@intvaluey<\r@p@sbblly
			      \xdef\r@p@sbblly{\number\p@intvaluey}\fi
			   \ifnum\number\p@intvaluey>\r@p@sbbury
			      \xdef\r@p@sbbury{\number\p@intvaluey}\fi
			   }
			\rotate@{\@p@sbbllx}{\@p@sbblly}
			\minmaxtest
			\rotate@{\@p@sbbllx}{\@p@sbbury}
			\minmaxtest
			\rotate@{\@p@sbburx}{\@p@sbblly}
			\minmaxtest
			\rotate@{\@p@sbburx}{\@p@sbbury}
			\minmaxtest
			\edef\@p@sbbllx{\r@p@sbbllx}\edef\@p@sbblly{\r@p@sbblly}
			\edef\@p@sbburx{\r@p@sbburx}\edef\@p@sbbury{\r@p@sbbury}
		\fi
		\count203=\@p@sbburx
		\count204=\@p@sbbury
		\advance\count203 by -\@p@sbbllx
		\advance\count204 by -\@p@sbblly
		\edef\@bbw{\number\count203}
		\edef\@bbh{\number\count204}
}
\def\in@hundreds#1#2#3{\count240=#2 \count241=#3
		     \count100=\count240	
		     \divide\count100 by \count241
		     \count101=\count100
		     \multiply\count101 by \count241
		     \advance\count240 by -\count101
		     \multiply\count240 by 10
		     \count101=\count240	
		     \divide\count101 by \count241
		     \count102=\count101
		     \multiply\count102 by \count241
		     \advance\count240 by -\count102
		     \multiply\count240 by 10
		     \count102=\count240	
		     \divide\count102 by \count241
		     \count200=#1\count205=0
		     \count201=\count200
			\multiply\count201 by \count100
		 	\advance\count205 by \count201
		     \count201=\count200
			\divide\count201 by 10
			\multiply\count201 by \count101
			\advance\count205 by \count201
		     \count201=\count200
			\divide\count201 by 100
			\multiply\count201 by \count102
			\advance\count205 by \count201
		     \edef\@result{\number\count205}
}
\def\compute@wfromh{
		\in@hundreds{\@p@sheight}{\@bbw}{\@bbh}
		\edef\@p@swidth{\@result}
}
\def\compute@hfromw{
	        \in@hundreds{\@p@swidth}{\@bbh}{\@bbw}
		\edef\@p@sheight{\@result}
}
\def\compute@handw{
		\if@height 
			\if@width
			\else
				\compute@wfromh
			\fi
		\else 
			\if@width
				\compute@hfromw
			\else
				\edef\@p@sheight{\@bbh}
				\edef\@p@swidth{\@bbw}
			\fi
		\fi
}
\def\compute@resv{
		\if@rheight \else \edef\@p@srheight{\@p@sheight} \fi
		\if@rwidth \else \edef\@p@srwidth{\@p@swidth} \fi
}
\def\compute@sizes{
	\compute@bb
	\if@scalefirst\if@angle
	\if@width
	   \in@hundreds{\@p@swidth}{\@bbw}{\ps@bbw}
	   \edef\@p@swidth{\@result}
	\fi
	\if@height
	   \in@hundreds{\@p@sheight}{\@bbh}{\ps@bbh}
	   \edef\@p@sheight{\@result}
	\fi
	\fi\fi
	\compute@handw
	\compute@resv}
\def\OzTeXSpecials{
	\special{empty.ps /@isp {true} def}
	\special{empty.ps \@p@swidth \space \@p@sheight \space
			\@p@sbbllx \space \@p@sbblly \space
			\@p@sbburx \space \@p@sbbury \space
			startTexFig \space }
	\if@clip{
		\if@verbose{
			\ps@typeout{(clip)}
		}\fi
		\special{empty.ps doclip \space }
	}\fi
	\if@angle{
		\if@verbose{
			\ps@typeout{(rotate)}
		}\fi
		\special {empty.ps \@p@sangle \space rotate \space} 
	}\fi
	\if@prologfile
	    \special{\@prologfileval \space } \fi
	\if@decmpr{
		\if@verbose{
			\ps@typeout{psfig: Compression not available
			in OzTeX version \space }
		}\fi
	}\else{
		\if@verbose{
			\ps@typeout{psfig: including \@p@sfile \space }
		}\fi
		\special{epsf=\ps@predir\@p@sfile \space }
	}\fi
	\if@postlogfile
	    \special{\@postlogfileval \space } \fi
	\special{empty.ps /@isp {false} def}
}
\def\DvipsSpecials{
	\special{ps::[begin] 	\@p@swidth \space \@p@sheight \space
			\@p@sbbllx \space \@p@sbblly \space
			\@p@sbburx \space \@p@sbbury \space
			startTexFig \space }
	\if@clip{
		\if@verbose{
			\ps@typeout{(clip)}
		}\fi
		\special{ps:: doclip \space }
	}\fi
	\if@angle
		\if@verbose{
			\ps@typeout{(clip)}
		}\fi
		\special {ps:: \@p@sangle \space rotate \space} 
	\fi
	\if@prologfile
	    \special{ps: plotfile \@prologfileval \space } \fi
	\if@decmpr{
		\if@verbose{
			\ps@typeout{psfig: including \@p@sfile.Z \space }
		}\fi
		\special{ps: plotfile "`zcat \@p@sfile.Z" \space }
	}\else{
		\if@verbose{
			\ps@typeout{psfig: including \@p@sfile \space }
		}\fi
		\special{ps: plotfile \@p@sfile \space }
	}\fi
	\if@postlogfile
	    \special{ps: plotfile \@postlogfileval \space } \fi
	\special{ps::[end] endTexFig \space }
}
\def\psfig#1{\vbox {
	\ps@init@parms
	\parse@ps@parms{#1}
	\compute@sizes
	\ifnum\@p@scost<\@psdraft{
		\PsfigSpecials 
		\vbox to \@p@srheight sp{
			\hbox to \@p@srwidth sp{
				\hss
			}
		\vss
		}
	}\else{
		\if@draftbox{		
			\hbox{\fbox{\vbox to \@p@srheight sp{
			\vss
			\hbox to \@p@srwidth sp{ \hss 
			 \hss }
			\vss
			}}}
		}\else{
			\vbox to \@p@srheight sp{
			\vss
			\hbox to \@p@srwidth sp{\hss}
			\vss
			}
		}\fi

	}\fi
}}
\psfigRestoreAt
\setDriver
\let\@=\LaTeXAtSign

\def\sun{$_\odot$}
\def\lapp{\lower2pt\hbox{$\buildrel {\scriptstyle <}
   \over {\scriptstyle\sim}$}}
\def\gapp{\lower2pt\hbox{$\buildrel {\scriptstyle >}
   \over {\scriptstyle\sim}$}}
\def\deriv#1#2{{{\rm d}#1\over {\rm d}#2}}
\def\msun{{\rm M}_{\odot}}
\def\rsun{{\rm R}_{\odot}}
\def\lsun{{\rm L}_{\odot}}
\def\mdot{\dot \rm M}
\def\mpy{{\rm M}_{\odot} {\rm yr}^{-1}} 

\newcommand{\be}{\begin{equation}}
\newcommand{\ee}{\end{equation}}
\newcommand{\bcen}{\begin{center}}
\newcommand{\ecen}{\end{center}}
\newcommand{\di}{\partial}
\newcommand{\bprime}{\mbox{$\beta^{\prime}$}}
\newcommand{\subsect}{\subsection}
\newcommand{\sect}{\section}
\newcommand{\Ang}{\mbox{\raisebox{1.7ex}{$\tiny\circ$}\hspace{-0.5em}A}}
\newcommand{\kms}{\mbox{ km~s$^{-1}$}}
\newcommand{\Mpc}{\mbox{M$_\odot$~pc$^{-3}$}}
\newcommand{\msolar}{\mbox{M$_\odot$}}
\newcommand{\lsolar}{\mbox{L$_\odot$}}
\newcommand{\kpc}{\mbox{kpc}}

\def\deg{\ifmmode ^{\circ}
         \else $^{\circ}$\fi}
\def\pdeg{\ifmmode $\setbox0=\hbox{$^{\circ}$}\rlap{\hskip.11\wd0 .}$^{\circ}
          \else \setbox0=\hbox{$^{\circ}$}\rlap{\hskip.11\wd0 .}$^{\circ}$\fi}
\def\arcs{\ifmmode {^{\scriptscriptstyle\prime\prime}}
          \else $^{\scriptscriptstyle\prime\prime}$\fi}
\def\arcm{\ifmmode {^{\scriptscriptstyle\prime}}
          \else $^{\scriptscriptstyle\prime}$\fi}
\newdimen\sa  \newdimen\sb
\def\parcs{\sa=.07em \sb=.03em
     \ifmmode $\rlap{.}$^{\scriptscriptstyle\prime\kern -\sb\prime}$\kern -\sa$
     \else \rlap{.}$^{\scriptscriptstyle\prime\kern -\sb\prime}$\kern -\sa\fi}
\def\parcm{\sa=.08em \sb=.03em
     \ifmmode $\rlap{.}\kern\sa$^{\scriptscriptstyle\prime}$\kern-\sb$
     \else \rlap{.}\kern\sa$^{\scriptscriptstyle\prime}$\kern-\sb\fi}
\def\gtorder{\mathrel{\raise.3ex\hbox{$>$}\mkern-14mu
             \lower0.6ex\hbox{$\sim$}}}
\def\ltorder{\mathrel{\raise.3ex\hbox{$<$}\mkern-14mu
             \lower0.6ex\hbox{$\sim$}}}

  \def\aa{{ A\&A}}
  \def\aj{{ AJ}}
  \def\annrev{{ ARA\&A}}
  \def\apj{{ ApJ}}
  \def\apjl{{ ApJ}}
  \def\apjs{{ ApJS}}
  \def\mn{{ MNRAS}}
  \def\Nature{{ Nature}}
  \def\science{{ Science}}
  \def\pasp{{ PASP}}
  \def\rmp{{ RevModPhys}}

\def\refindent{\par\penalty-100\noindent\parskip=4pt plus1pt
               \hangindent=3pc\hangafter=1\null}
\def\ref#1#2#3#4{\refindent#2, {#1\/,\ }{#3}, #4}
\def\reft#1#2#3#4#5#6{\refindent#2 {\it #6}, #3, {\it #1\/,\ }{\bf#4}, #5.}
\def\reftb#1#2#3#4#5#6{\refindent#2 #3, {\it #6}, {\it #1\/,\ }{\bf#4}, #5.}
\def\refbook#1{\refindent#1}
\def\preprint#1#2#3{\refindent#1, #2, {\it #3 preprint} }
\def\preprintt#1#2#3#4{\refindent#1, #2 {\it #4}, {\it #3 preprint}.}
\def\refinpress#1#2{\refindent#1, {\it #2, in press}.}
\def\reftinpress#1#2#3#4{\refindent#1 {#4}, #2, {\it #3, in press}}
\def\reftsubmit#1#2#3#4{\refindent#1 {#4}, #2, {\it #3, submitted}}
\def\bysame{\hbox to 50pt{\leaders\hrule height 2.4pt depth -2pt\hfill .\ }}

\def\sdml{$\sigma_{DM}$-$L$ }
\def\sdmlmu{$\sigma_{DM}$-$L$-$\mu$ }
\def\sdmlre{$\sigma_{DM}$-$L$-$R_e$ }

\begin{document}

\title{Shear \& Ellipticity in Gravitational Lenses }
\author{C.R. Keeton}
\author{C.S. Kochanek}
\author{U. Seljak}
\affil{Harvard-Smithsonian Center for Astrophysics, MS-51\protect \\
       60 Garden Street \protect \\
       Cambridge MA 02138 }
\authoremail{ckochanek@cfa.harvard.edu}

\begin{abstract}
Galaxies modeled as singular isothermal ellipsoids with an axis ratio distribution
similar to the observed axis ratio distribution of E and S0 galaxies are statistically
consistent with both the observed numbers of two-image and four-image lenses and
the inferred ellipticities of individual lenses.  
However, no four-image lens is well fit by the
model (typical $\chi^2/N_{dof} \sim 20$), the axis ratio of the model can
be significantly different from that of the observed lens galaxy, and the major
axes of the model and the galaxy may be slightly misaligned.  We found that
models with a second, independent, external shear axis could fit the data well (typical 
$\chi^2/N_{dof} \sim 1$), while adding the same number of extra parameters to the
radial mass distribution does not produce such a dramatic improvement in the fit.
An independent shear axis can be produced by misalignments between the luminous 
galaxy and its dark matter halo, or by external shear perturbations due to galaxies 
and clusters correlated with the primary lens or along the line of sight.  We estimate 
that the external shear perturbations have no significant effect on the expected 
numbers of two-image and four-image lenses, but that they can be important 
perturbations in individual lens models. However, the amplitudes of the external 
shears required to produce the good fits are larger than our estimates for typical external 
shear perturbations (10-15\% shear instead of 1-3\% shear) suggesting that the origin
of the extra angular structure must be intrinsic to the primary lens galaxy in
most cases.  
\end{abstract}


\section{Introduction}

We generally think of dark matter in galaxies in the context of the
radial distribution of mass.  Galaxy rotation curves for spirals, 
and X-ray halos (e.g. Fabbiano 1995) and gravitational lensing (Maoz \& Rix 1993, Kochanek 
1995, 1996a) for early type galaxies all require radial mass 
distributions that decline more slowly than the luminosity.  Stellar 
dynamical models of early-type galaxies can generally be made consistent 
with dark matter either present or absent depending on the assumptions 
about the structure of the stellar orbits (Saglia et al. 1993, Carollo et al. 1995). 
However, once we accept 
that the radial mass distribution is substantially composed of dark matter, 
there is little basis for believing that the ellipticity of the luminous 
matter is quantitatively representative of the ellipticity of the overall 
mass distribution.  In spiral galaxies we accept this premise as a matter 
of course -- the light distribution is a flattened disk, but the dark matter
distribution is a moderately oblate spheroid.  

The shapes of early-type galaxies are a mixture of oblate, prolate, and 
triaxial spheroids, although the inferred shape distributions (e.g. Schechter (1987),
Ryden (1992),  J\o rgensen \& Franx (1994)) are limited 
by the degeneracies inherent in deprojection (e.g. Rybicki 1987) and by 
triaxiality.  Kinematic misalignments between the projected angular 
momentum vector and the minor axis of the projected galaxy provide clear 
geometric evidence that the intrinsic shapes of ellipticals are triaxial 
(Binney 1985, Franx et al. 1991).  The observed axis ratios of the 
combined E \& S0 population show a deficit of round galaxies, a plateau
for axis ratios from 0.9 to 0.6, and a sharp decline beyond 0.5. 
Observational evidence on the shape of the mass distribution
in early type galaxies is very limited. Models of the rare 
polar-ring galaxies (e.g. Arnaboldi \& Sparke 1994, Sackett et al. 1994)
give inferred density axis ratios of 
$0.3 \ltorder c/a \ltorder 0.6$, similar or flatter than the inferred axis 
ratios of the central galaxy.  Buote \& Canizares (1994, 1995, 1996) studied 
the ellipticity of the X-ray halos of several elliptical galaxies.  In NGC 720 
they found that the dark matter had $0.3 \ltorder c/a \ltorder 0.5$, and in 
NGC 1332 they found $0.2 \ltorder c/a \ltorder 0.7$, and the
dark matter was at least as flattened as the luminous galaxy.  In NGC 720,
the projected major axes of the dark and luminous matter were misaligned by
$30^\circ \pm 15^\circ$ (90\% confidence).  
Theoretical models of galaxy formation predict ellipticities and triaxialities 
for the mass distribution larger than observed for luminous galaxies 
(Dubinski 1992, 1994, Warren et al. 1992).  Although the calculations 
contained only dark matter, some differences between the dark matter and
the baryons should persist as a consequence of the different dissipative 
processes in dark matter and gas.   

Gravitational lenses are sensitive to the angular structure of the
projected mass distribution of the lens galaxies through both the 
distributions of image morphologies (two-image, four-image etc.)
and detailed models of individual lenses.  We already know from
circular models of lenses that the radial mass distribution of 
successful lens models is inconsistent with constant mass-to-light
ratio dynamical models both from statistical studies (Maoz \& Rix
1993, Kochanek 1993, 1996a) and models of individual lenses
(e.g. Kochanek 1995, Grogin \& Narayan 1996).   Gravitational lenses supply three probes 
of the angular structure.  The first
is the ellipticity required to produce the observed 
numbers of two-image and four-image lenses.  The second 
is the ellipticity required to fit particular lensed systems,
and the third is the alignment of the inferred mass 
distribution with the observed lens galaxy.  If galaxies
contain dark matter, then the ellipticities of the light 
distributions need not match those of the mass distributions,
and if the dark matter distribution has either a different
triaxiality than the luminous distribution or is not in
dynamical equilibrium, the models need not be aligned with the
observed galaxy.  The numerical models of halo formation
predict that the mass distributions should be both more elliptical
and more triaxial than the light distribution.

Statistical models using non-circular lenses (Kochanek \& Blandford 1987, 
Kochanek 1991b, Wallington \& Narayan 1993, Kassiola \& Kovner 1993) focused 
on the existence of detectable numbers of bright four-image quasar lenses 
as a consequence of magnification bias, and they did not quantitatively 
evaluate the ellipticities required to fit the observed distribution of lens 
morphologies.  More recently, King et al. (1996) pointed out that the models 
used in these studies clearly produce fewer four-image lenses than are 
observed in the JVAS radio lens survey (Patnaik 1994, Patnaik et al. 1992).  
Kochanek (1996b) quantified the mismatch in greater detail for both the JVAS 
survey and optical quasar lens surveys.  Models of individual lenses strongly 
constrain the lens ellipticity for a given radial mass distribution, with 
centrally concentrated distributions requiring higher ellipticities than 
extended, dark matter distributions (see Kochanek 1991a, Wambsganss \& 
Paczy\'nski 1994).  Except for these preliminary surveys, there
are no systematic comparisons of the angular properties of lens models to the
expected properties of galaxies. 

There is, however, a complication to any program using lenses to study
the angular structures of galaxies -- the primary lens galaxy is not
the only source of shear in a gravitational lens.  The most important
sources of external shears are galaxies or clusters correlated with 
the primary lens galaxy (Kochanek \& Apostolakis 1988),
galaxies or clusters near the line of sight but at a different 
redshift (Kochanek \& Apostolakis 1988, Jaroszy\'nski 1991) and 
perturbations from large scale structure (Gunn 1967, Jaroszy\'nski et al. 
1990, Jaroszy\'nski 1991, Seljak 1994, Bar-Kana 1996).  Strong external
perturbations are rare, with only two lenses clearly requiring multiple
components in the lens model.  The lens 0957+561 (Walsh, Carswell \& Weymann 
1980) is a composite consisting of a galaxy and a cluster 
(Young et al. (1980), most recently modeled by Grogin \& Narayan 1996),
and the lens 2016+112 (Lawrence et al. 1984) has two lens galaxies 
that may be at different redshifts (e.g. Nair \& Garrett 1996). 
Weak external shear perturbations can also be probed by observing 
the correlations of galaxy ellipticities (e.g. Blandford et al. 1991,
Miralda-Escud\'e 1991, Kaiser 1992) or by measuring the weak shear
produced by individual galaxies (e.g. Valdes et al. 1984, Brainerd
et al. 1996).   

In our analysis we will quantitatively survey the origins of ellipticity and 
shear, the numbers of lenses of different morphologies, and models of 
the individual lenses and compare the results to the optical properties
of early type galaxies.  In \S2 we review the sources of shear in gravitational 
lenses: the primary lens galaxy, objects clustered with the lens galaxy, 
objects near the ray path, and shear from large scales structures.  In \S3 we 
briefly summarize the lens data we use in our analysis, and our analytical 
procedures.  In \S4 we consider the case of lensing by isolated early-type 
galaxies.  In \S5 we study the effects of adding additional sources of 
shear to the lens model.  Finally, in \S6 we review our results.

\section{The Sources Of Shear In Gravitational Lenses} 

The ellipticity and shear in a gravitational lens comes from three sources:  
the primary lens galaxy, galaxies or clusters near the lens galaxy, and 
structures along the ray path.  The primary galaxy is characterized by its 
ellipticity or axis ratio, and all external perturbations can be characterized 
by their external shear $\gamma$.  The structures along the ray path range 
from weak potential fluctuations to galaxies and clusters.  We first define the
general gravitational lensing equations needed to describe our models, and
then discuss each of the sources of ellipticity and shear.

We are interested in the case where there is already a strong lens, and
the effects of weak shear perturbations are to modify the lens equations into a
``generalized quadrupole lens'' (Kovner 1987). Relative to a fiducial ray
passing from the observer through the lens  center in the absence of the
lens potential, the lens equation, following the notation of Bar-Kana (1996), is
\begin{equation}
  \vec{u} = ( I + F_{OS}) \vec{x} -
        (I + F_{LS}) \vec{\alpha}\left[ ( I + F_{OL} ) \vec{x} \right]
\end{equation}
where $\vec{u}$ is the angular position of the source compared to the
fiducial ray, $\vec{x}$ is the angular position in the lens plane compared
to the fiducial ray, $\vec{\alpha}$ is the deflection produced
by the primary lens galaxy, and $I$ is the $2\times2$ identity matrix.
The $2\times2$ tensors $F_{OS}$, $F_{OL}$, and $F_{LS}$ describe 
additional shear and convergence due to perturbations between the
observer and the source, the observer and the lens, and the lens and
the source respectively.  Each tensor can be decomposed into a convergence 
$\kappa$, a shear $\gamma$, and orientation of the major axis of the shear
$\theta$, where $\kappa=(1/2)(F_{11}+F_{22})$,
$\gamma_c=\gamma\cos 2\theta = (1/2)(F_{11}-F_{22})$,
$\gamma_s=\gamma\sin 2\theta = F_{12} = F_{21}$, and
$\gamma^2= \gamma_c^2 + \gamma_s^2$. 

Statistical calculations are simplified by using the ``equivalent plane'' 
defined by the coordinates $\vec{X}= (1+F_{OL})\vec{x}$ and 
$\vec{U}=(1+F_{LS})^{-1}\vec{u}$ (Kovner 1987).  The lens equation in 
the equivalent plane is
\begin{equation}
\vec{U} = (I-F_e)\vec{X} - \vec{\alpha}(\vec{X}).
\end{equation}
In these coordinates, and to linear order, the effects of the
three shear tensors reduce to a single effective shear and convergence tensor, 
$ F_e = F_{OL} + F_{LS} - F_{OS}$, added to the effects of the primary lens 
(Bar-Kana 1996).  The advantage of the
effective plane is that cross sections and magnification
probability distributions depend only on $F_e$ in the effective plane but
are easily transformed back into the normal coordinates.
If $\sigma'$ is a cross section computed in the equivalent plane, then the
cross section in the original coordinates is
$\sigma' | I + F_{LS} |^{-1}$, and if $M'$ is a
magnification computed in the equivalent plane, then the magnification
in the original coordinates is $M^{-1}= (M')^{-1} |I+F_{LS}||I+F_{OL}|$,
where $|\cdots|$ denotes a determinant.
If the convergence and shear of $F$ are $\kappa$ and $\gamma$,
then $|I+F|=(1-\kappa)^2 - \gamma^2$.  If we only use cross sections and
magnifications computed in the effective plane, we make errors  
in the statistical calculation that are first order in $\kappa$
(second order if $\langle \kappa \rangle=0$ as in the LSS model of \S2.3)
and second order in $\gamma$.

\subsection{The Primary Lens Galaxy}

The primary lens galaxy in most lens systems is an early type galaxy (E or S0),
with only 10--20\% of lenses contributed by spiral galaxies (Fukugita \& Turner
1991, Maoz \& Rix 1993, Kochanek 1996).  Given the general mass uncertainties
for galaxies, we can lump the E and S0 galaxies into a common population for
the purposes of lens models.  For simplicity 
the calculations will neglect the effects of misalignments 
between the luminous galaxy and the dark matter halo, although we 
consider whether the problems in the model can be explained by these effects.  
Although we are assuming a dark matter model, we would like to compare
our inferences to the observed axis ratios of luminous galaxies.
We used the J\o rgensen \& Franx (1994)
sample of 53 E and 93 S0 galaxies in the Coma cluster to estimate the 
axis ratio distribution of E and S0 galaxies.  The natural ellipticity 
parameter 
for lens models is the eccentricity $\epsilon$, where the two-dimensional 
axis ratio is $q_2^2=(1-\epsilon)/(1+\epsilon)$. We modeled the eccentricity
distribution with a Gaussian, 
$dP/d\epsilon \propto \exp(-(\epsilon-\epsilon_0)^2/2\Delta\epsilon_0^2)$
($0 \leq \epsilon \leq 1$),
and the peak Kolmogorov-Smirnov (K--S) test probability for fitting the joint E+S0 
distribution was 94\% for parameters of $\epsilon_0 = 0.26$ and $\Delta \epsilon_0 = 0.33$.  
The E galaxies have lower mean ellipticities (peak of 82\% for $\epsilon_0=0.14$ 
and $\Delta \epsilon_0=0.15$) than the S0 galaxies (peak of 94\% for 
$\epsilon_0=0.44$ and $\Delta\epsilon_0=0.21$).  The mean ellipticity of the
ellipticals is underestimated because many low ellipticity S0 galaxies are 
misclassified as ellipticals (see J\o rgensen \& Franx 1994).  If the ellipticity of the 
light is a guide to the ellipticity of the mass, high ellipticity lens 
galaxies are S0 galaxies and low ellipticity lens galaxies are ellipticals.

We limited our study to a single model for the primary lens galaxy, the singular
isothermal ellipsoid.  We chose the model because singular isothermal spheres (SIS) are
the only models known to be simultaneously consistent with gravitational lens statistics, 
models, and stellar dynamics (see Kochanek 1996a).  We consider other monopole structures 
in Keeton \& Kochanek (1996b).  The singular isothermal ellipsoid was 
used by  Kassiola \& Kovner (1993) to study the statistics of lensed quasars and by 
Kormann et al. (1994a) to model B 1422+231.  Kassiola \& Kovner (1993) and Kormann et al. (1994b) 
discuss the model's analytic 
properties in detail.  The projected surface density of the singular isothermal ellipsoid is 
\begin{equation}
    2 { \Sigma \over \Sigma_c} = { b \over r }
       {\eta (\epsilon) \over (1-\epsilon \cos 2\theta)^{1/2} }
\end{equation}
where $\Sigma_c =c^2 (1+z_l)D_{OS}/4\pi G D_{OL} D_{LS}$ is the critical surface density for 
gravitational lensing, and $b=4\pi(\sigma/c)^2 D_{LS}/D_{OS}$ is the tangential critical
radius of the circular SIS for a one-dimensional dark matter velocity dispersion
of $\sigma$ normalized to match the observed line-of-sight velocity dispersions 
$\langle v_{los}^2\rangle$ of galaxies (see Kochanek 1994).   We use only an 
$\Omega_0=1$ cosmological model, where the comoving distances are 
$D_{ij} = 2 r_H ((1+z_i)^{-1/2}-(1+z_j)^{-1/2})$ 
for $r_H=c/H_0$.  For $z_i=0$, $D_{ij}$ is the proper motion distance to $z_j$.  
The factor $\eta(\epsilon)$ is an ellipticity dependent normalization factor, and 
$\eta(\epsilon=0)\equiv 1$.

\begin{figure}
\centerline{\psfig{figure=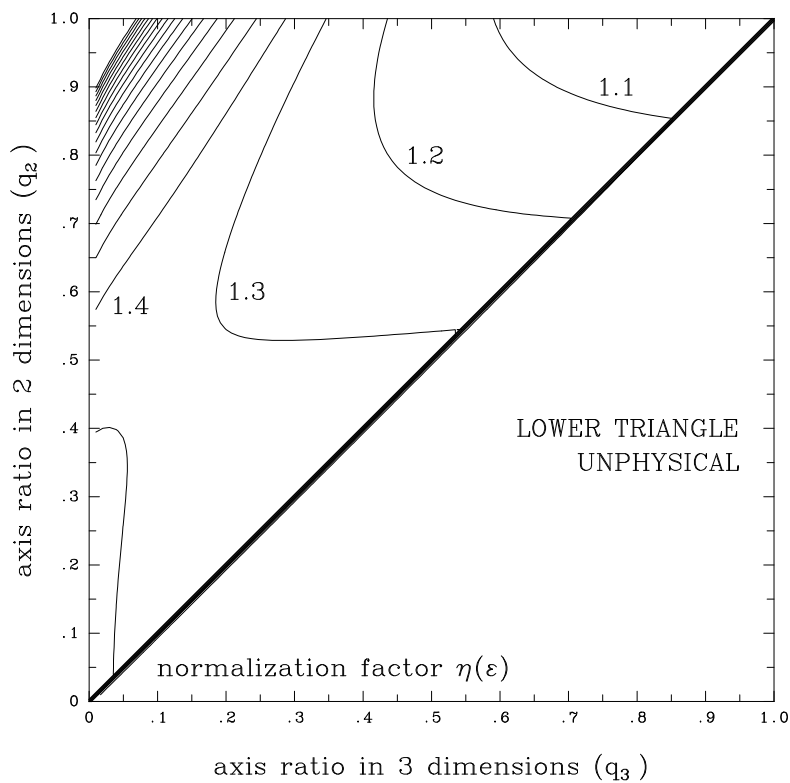,width=4.0in}}
\caption{The normalization factor $\eta$ for isothermal ellipsoids with 
 three-dimensional axis ratio $q_3$ and two-dimensional axis ratio $q_2$.  
 Only the upper-left triangle is physical because $q_2 \geq q_3$.  Contours 
 are spaced every 0.1 from $\eta=1$ for a spherical system to $\eta=3$ for 
 nearly pole-on ($q_2 \sim 1$) flattened systems ($q_3\sim0$).}
\end{figure}

Lens calculations are normalized to group lenses with fixed line-of-sight stellar
velocity dispersions, $\langle v_{los}^2 \rangle^{1/2}$,
so the normalization factor $\eta(\epsilon)$ should be defined so that 
$\langle v_{los}^2 \rangle$ is fixed as we vary $\epsilon$.  The stellar dynamics of
ellipsoidal stellar distributions in ellipsoidal dark matter distributions is a little
studied problem (e.g. de Bruijne, van der Marel \& de Zeeuw 1996), and a detailed examination 
is well beyond the
scope of our study.  To gain a qualitative
understanding of $\eta$, we consider the idealized problem in which both
the luminosity and density distributions are axisymmetric, oblate ($0<q_3 < 1$)
singular isothermal ellipsoids with density
\begin{equation}
  \rho = { \sigma_0^2 \over 2 \pi G } { 1 \over R^2 + z^2/q_3^2 }.
\end{equation}
and projected surface density 
\begin{equation}
   \Sigma =   { \sigma_0^2 (1+\epsilon) q_3 \over
                2 G R (1-\epsilon \cos 2\theta)^{1/2} }
   \quad\hbox{where}\quad
   \epsilon = { 2 (1-q_3^2) \sin^2 i \over 3 + q_3^2 + (1-q_3^2) \cos 2 i },
\end{equation}
and $i$ is the inclination angle of the galaxy relative to the observer
($i=0$ is pole on).  The velocity $\sigma_0$ is an
unmeasurable parameter that must be related to the line-of-sight
velocity dispersion of the stars, $\langle v_{los}^2 \rangle$, through
the normalization factor $\eta = \sigma_0^2 (1+\epsilon) q_3/ \langle v_{los}^2 \rangle$.

We can determine the velocity dispersions of the isothermal ellipsoid analytically 
for an axisymmetric two-integral dynamical model (Binney \& Tremaine 1987, \S4.2) assuming
a constant mass-to-light ratio with stellar density $\nu=\rho$.
The velocity dispersions in the cylindrical $R$ and $Z$ directions are equal
with
\begin{equation}
  \nu\sigma^2_R = \nu\sigma^2_Z = { \nu_0 \sigma_0^2 \over R^2 a^2 }
      \left[ \left(\tan^{-1} a \right)^2 -
             \left(\tan^{-1} a |x| \right)^2 \right]
\end{equation}
where $a^2=e^2/(1-e^2)$, $e^2=1-q_3^2$, and $r$ and $x=\cos\theta$ are spherical 
polar coordinates.  The mean square velocity in the direction of the cylindrical 
angle $\phi$ is
\begin{equation}
  \nu \langle v_\phi^2 \rangle =
        { 2 \sigma_0^2 \nu_0 \tan^{-1} a \over r^2 a (1+a^2 x^2) }
         - \nu \sigma^2_R.
\end{equation}
For fixed $\sigma_0$ in eqn. (4), adding ellipticity reduces the stellar velocity
dispersions relative to the spherical model at all positions in the galaxy. 
The line-of-sight velocity dispersion through
an infinite aperture can be reduced to a simple one-dimensional integral,
\begin{equation}
   { \langle v_{los}^2 \rangle \over \sigma_0^2 } =
    \sin^2 i { \tan^{-1} a \over a } +
    \cos^2 i \int_0^{\pi/2} dx {
    \left(\tan^{-1} a \right)^2 - \left(\tan^{-1} a x\right)^2
    \over a (1-x^2) \tan^{-1} a }.
\end{equation}
Figure 1 shows contours of $\eta$ as a function of the two- and three-dimensional 
axis ratios of the ellipsoid.  In general, the lensing surface density in eqn. (3) 
with $\eta=1$ underestimates lensing by flattened galaxies for a fixed line-of-sight
velocity dispersion at all inclination angles.  Edge-on, modestly flattened galaxies
($q_3 > 1/2$, $q_2 \sim q_3$) produce more lenses than pole-on or rounder galaxies,
so the normalization factor enhances the number of lenses produced by the more
elliptical lenses (e.g. Subramanian \& Cowling 1986). 
We will use $\eta=1$ for the lensing calculations, because our dynamical model is not
quantitatively accurate. With $\eta=1$ we underestimate the numbers of lenses produced by
high ellipticity galaxies, causing us to overestimate the numbers of 
high ellipticity galaxies required to fit the data by $\eta^2(\epsilon)$.

\subsection{Galaxies and Clusters Near the Primary Lens}

The galaxy correlation function enhances the probability of finding another 
galaxy near the primary lens, and it is more likely that a perturbing 
galaxy is at the redshift of the primary lens galaxy than elsewhere along the 
line of sight (see Kochanek \& Apostolakis 1988).  Early type 
galaxies also tend to live in high-density environments (e.g. Postman \& Geller 1984), further 
increasing the probability of finding a nearby perturbing galaxy.  We
consider the shear produced by the nearest neighbor galaxy, and the shear
produced by clusters of galaxies.  The shear from perturbers at the same
redshift as the primary lens have $F_{OS}$ non-zero, and $F_{LS}\equiv0$ and
$F_{OL}\equiv0$ in the generalized quadrupole lens of eqn. (1).  

In three dimensions, the galaxy-galaxy correlation function is 
$\xi(r) = (r/r_0)^{-\chi}$ where the comoving correlation length is
$r_0 \simeq 5 h^{-1}$ Mpc and the exponent is $\chi \simeq 1.75$ 
(e.g. Peebles 1995).  We model the shear from a neighbor by
$\gamma = \gamma_{bk} ( 2 a/r -1)$ for $r<a$ and $\gamma_{bk} a^2/r^2$
for $r>a$ where $\gamma_{bk} = b/4a$ is the shear at the break radius
$a \sim$ 50--200 kpc.  For $r <a $ the model becomes an SIS lens, while for $r>a$ it
is a point mass lens.  
If $\gamma > \gamma_{max} = 2b/r_{min} \simeq 1/4$, then the
two lenses have interacting caustics and the perturbation cannot be modeled
by a shear tensor (see Kochanek \& Apostolakis 1988). 
The optical depth for the nearest neighbor to produce a shear exceeding
$\gamma$ is
\begin{equation}
  \tau(>\gamma) \simeq 
     0.01  \left[ { n_* \over 10^{-2} h^3 \hbox{ Mpc}^{-3} }\right]
       \left[ { (1+z_l) r_0 \over 5 h^{-1} \hbox{Mpc} } \right]^{7/4}
       \left[ { \sigma_* \over 220 \kms } \right]^{5/2}
    \left[ { D_{OS} \over r_H } \right]^{5/4} x^{5/4} (1-x)^{5/4} \gamma^{-5/4}.
\end{equation}
for $\chi=1.75$, and a constant comoving density of lenses $n_*$ with a Schechter 
(1976) function exponent $\alpha=-1$ and a ``Faber-Jackson'' exponent 
$\gamma_{FJ}=4$ (see \S3), where $x=D_{OL}/D_{OS}$ is the fractional distance 
of the primary lens from the observer.  The ratio $D_{OS}/r_H=1$ at 
$z_s\simeq 3$ for an $\Omega_0=1$ cosmological model.  
The peak shear perturbation is found at $x=1/2$, the optimal distance for the
primary lens, where the optical depth reaches 
$\tau(>\gamma) \simeq 0.002(D_{OS}/r_H)^{5/4} \gamma^{-5/4}$.
These expressions fail when the optical depth approaches unity (typically
at impact parameters smaller than $a$, which allows us to ignore the regime $r>a$).  
We can approximate the
minimum shear scale by truncating the optical depth at $\tau(>\gamma_{min})=1$, where 
$\gamma_{min} \simeq 0.025 x (1-x) (D_{OS}/r_H)\ltorder 0.005(D_{OS}/r_H)$.  
The optical depth for non-linear interactions between galaxies is 
\begin{equation}
 \tau(\gamma>\gamma_{max}\simeq 1/4) 
    \simeq 0.06 x^{5/4} (1-x)^{5/4} \left( { D_{OS} \over r_H } \right)^{5/4} 
    \ltorder 0.01 \left( { D_{OS} \over r_H } \right)^{5/4}
\end{equation}
so lenses with more than one primary lens galaxy are rare (as observed).

The opposite limit to considering only the nearest neighbor galaxy
is to put the primary lens galaxy in a cluster.  The local comoving density of
clusters as a function of velocity dispersion $\sigma$ is approximately 
$dn/d\sigma \simeq (n_0/\sigma_0)(\sigma/\sigma_0)^{-c}$ with
$c \simeq 8.4\pm 1.0$, $n_0=3.6^{+1.6}_{-1.0} \times 10^{-3} h^3$ Mpc$^{-3}$ 
and $\sigma_0 = 400 \kms$ using Henry \& Arnauld's (1991) X-ray luminosity 
function and assuming $\sigma^2 = k T / \mu m_p$ ($\beta \simeq 1$) to relate 
the velocity dispersion $\sigma$ to the X-ray temperature $T$.  Let the 
cluster mass distribution be an SIS truncated at outer radius $r_c$ where
$r_c = r_{c0} (\sigma/\sigma_0)^y$, and the galaxy distribution 
in the cluster be $n_g (r_g/r)^2$ with the same outer radius.
The cluster shear is $\gamma =b/2 r$ where $b=b_0(\sigma/\sigma_0)^2$,
$b_0 = 4\pi (\sigma_0/c)^2 D_{LS}/D_{OS}$, the shear at
the cluster edge is $\gamma_{min}=\gamma_0 (\sigma/\sigma_0)^{2-y}$,
and $\gamma_0 = b_0/2r_{c0}$.  The shear probability distribution
is convergent and dominated by the low mass clusters if $c > 3 +y$. 
We now consider only the self-similar solution with $y=2$ for which the shear 
distribution is independent of the cluster velocity dispersion.  The integral 
optical depth is
\begin{eqnarray}
  \tau(>\gamma) 
     &= &{ 8 \pi^2 \over c- 5} n_0 r_H [(1+z_l)r_{c0}]^2 \left( { \sigma_0\over c} \right)^2
                  { D_{OS} \over r_H} { x(1-x) \over \gamma} \nonumber \\
     &= &0.01\left[ { n_0 \over 4 \times 10^{-3} h^3 \hbox{Mpc}^{-3} } \right]
                        \left[ { (1+z_l) r_{c0} \over 5 h^{-1} \hbox{Mpc} } \right]^2
                        \left[ { \sigma_0 \over 400 \beta^{1/2} \kms } \right]^2
                        { D_{OS} \over r_H} { x(1-x) \over \gamma}
\end{eqnarray}
assuming a constant comoving density of clusters truncated at $\sigma_0$.  
Thus if every lens galaxy is in a cluster, the shear contribution from the 
clusters is nearly equal to the shear from the nearest neighbor galaxies 
(eqn. 10).  However, by assuming a constant comoving density of clusters, 
eqn. (12) significantly overestimates the contribution from clusters, so we 
expect that the shear distribution is dominated by the nearest neighbor 
galaxies.

{\it The cluster shear contribution is dominated by the groups and small
clusters with a negligible contribution from large clusters.}  
Since the fraction of the optical depth from clusters with velocity 
dispersions exceeding $\sigma$ is $(\sigma/\sigma_0)^{-3.4\pm1.0}$, half of 
the optical depth is from velocity dispersions within 20\% of the lower limit 
(between $\sigma_0$ and $1.2\sigma_0$). In the observed lens sample it is
clear that large clusters are unimportant because lensing by rich clusters 
is found only by looking at rich clusters as part of surveys for arcs (see 
Kneib \& Soucail 1996).  The only convincing cluster-scale lens found by 
searching for lensed sources (i.e. not selected based on the mass of the lens)
is 0957+561 (Walsh et al. 1979, Young et al. 1980), where the cluster is 
exactly the type of poor, sparse cluster expected to dominate the statistics.  
The convergence produced by a cluster is tightly correlated with the shear 
($\kappa=\gamma$ for the SIS model), so the optical depth for large 
convergences is also small.  Previous models of multiply imaged
quasars considered the extra convergence produced by clusters, but not
the extra shear (e.g. Turner, Ostriker, \& Gott 1984, Maoz \& Rix 1993, 
Kochanek 1993).   Since large shear perturbations are easily detected,
it is unlikely that cluster convergences can be significantly distorting
gravitational lens statistics.

\subsection{Perturbations Along the Line-of-Sight: Large Scale Structure}

We next consider the shear generated by weak, long-wavelength inhomogeneities 
along the line of sight, which we will refer to as large scale structure (LSS)
shear, following the approach of Bar-Kana (1996) based on earlier studies by 
Gunn (1967), Kovner (1987), Kaiser (1992) and Seljak (1994, 1996). Light 
propagating through the universe is deflected by inhomogeneities along the line of 
sight. Although the total deflection angle can be of the order of arcminutes,
the relative distortion of a bundle of photons is small.  Such weak lensing
does not generate multiple images, but it does distort the shapes of 
background sources.  The distortions may be measurable from the 
ellipticities of high redshift galaxies (e.g. Blandford et al. 1991; 
Miralda-Escud\'e 1991; Kaiser 1992).  

Potential fluctuations between the observer and the primary lens produce
$F_{OS}$ and $F_{OL}$ shear terms in the generalized quadrupole lens (eqn. (1)),
while potential fluctuations between the primary lens and the source
produce $F_{OS}$ and $F_{LS}$ terms.  The $F_{LS}$ term is observable 
only in the lens time delays, so we focus on the effective shear 
$F_e = F_{OL} + F_{LS} - F_{OS}$ and the $F_{OL}$ shear.   
If we assume linear evolution of the power spectrum in an $\Omega_0=1$
universe, then the power spectrum of the potential fluctuations, 
$P_\phi(k,z) = 9 \Omega^2 H^4 a^4 \Delta^2 (k)/16\pi k^7$, is
independent of redshift, and ensemble averages of the LSS shear terms depend only on
\begin{equation}
   G =  r_H^3 \int_0^\infty k^5 dk P_\phi(k) =
     { \Omega_0^2 \sigma_R^2 \over 16 \pi r_H k_0 } 
 { \int_0^\infty dq q^{n+1} T^2(q) \over
   \int_0^\infty dq q^{n+2} T^2(q) [\sin(u)-u\cos(u)]^2/u^6 }.
\end{equation}
The power spectrum is defined by $\Delta^2(k) = A k^{n+3} T_k(k)^2$ 
where $A$ is a normalization constant, $k^n$ is the primordial spectrum 
($n\simeq1$),
\begin{equation}
  T_k = { \ln (1+2.34 q) \over q } \left[
  1 + 3.89 q + (14.1 q)^2 + (5.46 q)^3 + (6.71 q)^4 \right]^{-1/4}
\label{BBKS}
\end{equation}
is the BBKS (Bardeen et al. 1986) transfer function where $q=k/k_0$ and
$k_0= \Omega_0 h^2 $ Mpc$^{-1}$, $u= R k = R k_0 q$, and the mean square density
fluctuation in a top-hat filter of radius $R$ is
\begin{equation}
   \sigma_R^2 = \int_0^\infty  \Delta^2(k) { dk \over k}
     { 9 \over (k R)^6 } \left( \sin k R - k R \cos k R \right)^2.
\end{equation}
For $\Omega_0 h = 0.25$ and $n=1$, we find 
$G = 1.2 \times 10^{-4}\sigma_8^2 h^{-1}$.  Eqn. (13) is strictly valid
only for $\Omega_0=1$ and linear evolution, but the results found
using non-linear power spectrum models (Jain, Mo \& White (1995),
Peacock \& Dodds (1996)) are relatively accurate (10--20\% errors)
over a wide range of $\Omega_0$ and $\lambda_0$.   While the gravitational
potential in $\Omega_0 < 1$ models decays in a linear model, the
non-linear evolution and the longer path-lengths nearly cancel 
the decay so that eqn. (13) is roughly correct for other 
cosmological models. 

The mean square effective shear, OL shear, and their correlation are determined by
the weighted average over the potential $G$ and geometry. We find that
\begin{eqnarray}
  \langle \gamma_{e}^2 \rangle 
       &= &{ 4 \pi^2 \over 5} G \left( { D_{OS} \over r_H }\right)^3 x^2(1-x)^2 \nonumber \\
  \langle \gamma_{OL}^2 \rangle 
       &= &{ 2 \pi^2 \over 15} G \left( { D_{OS} \over r_H }\right)^3 x^3  \\
  \langle \gamma_{OL}\gamma_e \rangle 
       &= &-{  \pi^2 \over 5} G \left( { D_{OS} \over r_H }\right)^3 x^3 (1-x). \nonumber 
\end{eqnarray}
The maximum effective shear $\gamma_{em}^2 = (\pi^2 G /20)(D_{OS}/r_H)^3$ is found for a 
primary lens at one-half the distance to the source $x=D_{OL}/D_{OS}=1/2$ and it scales 
with the source distance as $D_{OS}^{3/2}$ (Bar-Kana 1996).  The deformation of the 
lens plane $\gamma_{OL}$ is determined by 
the $F_{OL}$ matrix, and it is strongest for a primary lens near the source.
The $\gamma_e$ and $\gamma_{OL}$ shears are weakly anti-correlated, 
with a mean angle $\Delta \theta$ between them of $\langle \cos \Delta \theta \rangle = -(3x/8)^{1/2}$.  
We can more usefully characterize the average properties of the LSS shear
by averaging over the lens cross section.  For a SIS lens in a flat
cosmology the differential optical depth is $d\tau \propto D_{OL}^2 D_{LS}^2 d D_{OL}$,
and the cross section averaged shears are
$\langle \gamma_e^2\rangle = (16/21) \gamma_{em}^2$,
$\langle \gamma_{OL}^2 \rangle = (10/21) \gamma_{em}^2$ and
$\langle \cos \Delta \theta\rangle = - (5/32)^{1/2}$.

Figure 2 summarizes the strength of the LSS shear including the effects of 
non-linear power-spectra, the cosmological model, and $\sigma_8$.  In the 
linear case the maximum rms LSS shear $\gamma_{em}$ is only 1--2\% for 
$\sigma_8 \simeq 0.5$ (Eke et al. 1996) and a source at $z_s \sim 3$. The 
dominant contribution comes from Mpc scales, corresponding to the outer 
parts of clusters and to superclusters.  In the non-linear case, the maximum 
rms shear is 4--6\% and the dominant contribution comes from linear scales 
near $k^{-1} \sim 100$ kpc.  In all models the strength of the LSS shear
increases with $\sigma_8$ and $\Omega_0 h$. 

\begin{figure}
\centerline{\psfig{figure=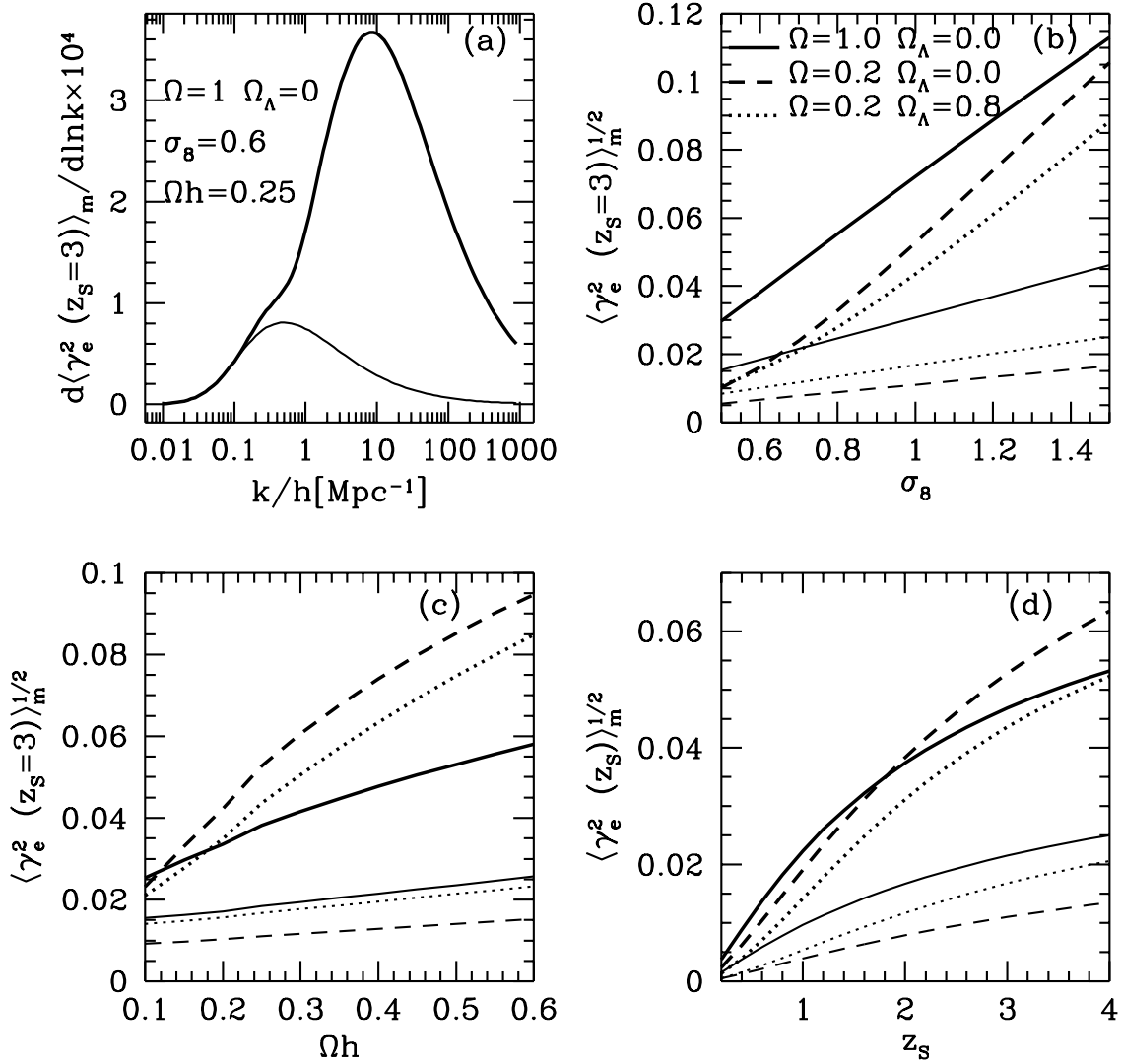,width=5.0in}}
\caption{ The dependence of LSS shear on cosmology for both linear (light)
and non-linear (dark) power spectrum models.  Panel (a) shows 
the logarithmic contribution to $\gamma_m(z_S=3)$
as a function of wavevector $k$ for linear (light solid) and non-linear
(heavy solid) power spectrum models. Panel (b) shows the dependence of
$\gamma_m(z_S=3)$ on the amplitude $\sigma_8$ 
for a fixed shape $\Omega_0 h = 0.25$. Panel (c) shows the dependence 
on the shape $\Omega_0 h$ for fixed amplitude, with $\sigma_8=0.6$ for
$\Omega_0=1$ and $\sigma_8=1.0$ for $\Omega_0 < 1$.  Finally, panel
(d) shows the variation of the shear with source redshift for 
the models in panel (c) fixed to $\Omega_0 h =0.25$.}
\end{figure}

\subsection{Perturbations Along the Line-of-Sight: Nonlinear Objects}

The LSS shear contribution is dominated by non-linear structures, and the
linear scale of $k^{-1}\sim 100$ kpc dominating the shear in the non-linear
power spectrum corresponds to a non-linear scale near $k^{-1} \sim 20$ kpc.
Hence the shear in the non-linear power spectrum estimates is not 
caused by ``large scale structure,'' but is an approximation
to the shear produced by collapsed halos.  Shear from galaxies along
the line of sight was considered by Kochanek \& Apostolakis (1988),
Jaroszynski (1991) and it is related to the efforts by Valdes et al. 
(1984) and Brainerd et al. (1996) to measure weak shears produced by 
individual galaxies. In this subsection we develop the effects of 
discrete halos near the line of sight using the language of the LSS 
shear calculation.

A single perturber of the primary lens 
produces $F_{OS}$ and $F_{OL}$ shear terms if it is in the foreground
and $F_{OS}$ and $F_{LS}$ shear terms if it is in the background.
For a lens at fractional distance $x=D_{OL}/D_{OS}$ to the source, the 
integral optical depth for $\gamma_e$ using an SIS perturber model is
\begin{eqnarray}  
   \tau(>\gamma_e) &= &{ 4 \over 5 } \pi^3 n_* r_H^3 
       \left( { \sigma_* \over c } \right)^4
       \left( { D_{OS} \over r_H } \right)^3 { x^2 (1-x)^2 \over \gamma_e^2 } \nonumber \\
     &= &0.002  \left[ { n_* \over 10^{-2} h^3 \hbox{ Mpc}^{-3} } \right]
              \left[ { \sigma_* \over 220 \kms } \right]^4
           \left( { D_{OS} \over r_H } \right)^3 { x^2 (1-x)^2 \over \gamma_e^2 }
\end{eqnarray}
and the integral optical depth for $\gamma_{OL}$ is
\begin{equation}
   \tau(>\gamma_{OL}) = { 2 \over 15 } \pi^3 n_* r_H^3 
          \left( { \sigma_* \over c } \right)^4
           \left( { D_{OS} \over r_H } \right)^3 { x^3 \over \gamma_{OL}^2 }.
\end{equation}
We can truncate the distribution either at the point where the
universe becomes optically thick, or at the value of the shear at the 
characteristic radius $a$ where the galaxy mass distribution begins to 
decline faster than the isothermal profile.  Generally these scales are
similar.  Comparing the optical depth
for $\gamma_e$ to the optical depth for correlated galaxies (eqn. (10)),
we find that the correlated term is the more important for shears
larger than $\gamma \gtorder 0.12 x(1-x)(D_{OS}/r_H)^{7/3} = 0.03(D_{OS}/r_H)^{7/3}$ for $x=1/2$.
The optical depths in eqns. (16) and (17) have the same distance scalings as the LSS shear
results in eqn. (15), so we can define an effective $G$ parameter (eqn. (12)) by 
\begin{equation}
   G = 2 \pi n_* r_H^3 \left( {\sigma_* \over c} \right)^4 \ln \Lambda 
     \simeq 5 \times 10^{-4} 
            \left[ { n_* \over 10^{-2} h^3 \hbox{ Mpc}^{-3} } \right]
            \left[ { \sigma_* \over 220 \kms } \right]^4
            \ln \Lambda.
\end{equation}
The ``Coulomb logarithm'' is $\ln \Lambda=\ln(\gamma_{max}/\gamma_{min}) \simeq 3.2$,
where $\gamma_{max} \simeq 0.25$ is the shear at which the caustics merge, and 
$\gamma_{min} \sim 0.01$ is the shear at which the universe becomes 
optically thick.  The magnitude of $G$ calculated using discrete non-linear
potentials qualitatively agrees with that calculated using the 
non-linear power spectrum in the LSS model ($G \simeq 3 \times 10^{-4}$). 

For weak shears the description in terms of discrete galaxies must fail 
because the universe is optically thick, so the true probability distribution 
for weak shears should approach the form expected for Gaussian random fields, 
$d\tau/d\gamma_e \propto \gamma_e \exp(-\gamma_e^2/\langle\gamma_e^2\rangle)$.  For 
strong shears, the universe is optically thin, and the distribution
must approach the discrete galaxy result, $d\tau/d\gamma_e \propto \gamma_e^{-3}$.
We explored how the two limits merge using Monte Carlo simulations of
lensing by a constant comoving density of Poisson-distributed galaxies.
As expected, for $\gamma_e \gtorder 0.1$ the shear is entirely due to
the nearest galaxy and has the power law distribution, while for 
$\gamma_e \ltorder 0.02$ it is due to random combinations of weak shears  
from many galaxies and the distribution approaches the Gaussian limit.
In the intermediate range, there is usually a dominant perturbing galaxy,
but the effects of other nearby galaxies cannot be neglected.

Since large shear perturbations are dominated by a particular galaxy near the
line of sight, we can estimate its properties. The mean distance
to a $\gamma_e$ perturber is the distance to the primary lens, $D_{OL}$,
while the mean distance to a $\gamma_{OL}$ perturber is one-half the
distance to the primary lens, $D_{OL}/2$.  The typical angular
separation of a $\gamma_e$ perturber from the primary lens is
$\simeq 0\parcs25/\gamma_e$, while the typical $\gamma_{OL}$ perturber
is further away in angle, $\simeq 0\parcs37/\gamma_{OL}$, because it is
physically closer to the observer.  A $\gamma_{OL}$ perturber is
generally brighter than the primary lens galaxy, both because it
is closer to the observer and because it is rarely ideally placed
as a lens (forcing it to be a more massive galaxy), while a $\gamma_e$
perturber is generally of comparable brightness to the primary
lens galaxy.  Finally, we can estimate the non-linearity of the perturber, 
or the fractional importance of the next order terms in the expansion of
the perturber's gravitational field beyond the shear tensor.  For $\gamma_e$
perturbers the non-linearity is of order $2\gamma_e \Delta\theta$
where $\Delta \theta$ is the diameter of the image system of the lens 
in arcseconds.  The $\gamma_{OL}$ perturbers are generally more non-linear,
with an average non-linearity of $6\gamma_{OL} \Delta\theta$.  
For a $\gamma_{OL}=0.05$ perturbation of a $\Delta\theta=2''$ lens, 
the next order terms have 64\% of the strength of the linear shear terms.

\section{Lens Data and Calculation Methods}

We confine our analysis to gravitational lenses in well-defined samples for which we
can make both models of the lenses and statistical models
of the number of two- and four-image lenses.   We use the
the data from the quasar surveys (Maoz et al. 1993ab, Crampton et al. 1992
Yee et al. 1993, Surdej et al. 1993, Kochanek, Falco \& Schild 1995) and the JVAS 
radio survey (Patnaik 1994, Patnaik et al. 1992, King et al. 1996).
We do not include
serendipitously discovered lenses, the MIT-Greenbank (MG, Burke et al. 1992)
or the CLASS survey (Jackson et al. 1995, Myers et al. 1995, Myers 1996)  
lenses primarily because we cannot make statistical
models of the numbers of lenses in these surveys.

We use the quasar lens sample used by Kochanek (1996a).  The sample contains 
three two-image lenses (0142--100, LBQS 1009--0252 and 1208--1011)  
and two four-image lenses (PG 1115+080, and H 1413+117).
We use the compact source subsample of the JVAS radio survey
which contains two two-image lenses (B 0218+357 and
a second double) and two four-image lenses (B 1422+231 and MG 0414+0534).  We exclude
B 1938+666 both because it has an extended source, which invalidates our 
statistical models, and because it has never been modeled.  
Table 1 lists the lenses we discuss and their properties.  Keeton \& Kochanek
(1996a) provides a more extensive summary of the data, and we will discuss
detailed models in Keeton \& Kochanek (1996b).

\begin{table}[t]
\begin{center}
\begin{tabular}{rccccc}
\multicolumn{6}{c}{Table 1: Lens Data} \\
\hline
\multicolumn{1}{c}{Lens}           &$z_s$  &$z_l$ &$N_{im}$ &$q_l$         &P.A.  \\
\hline
PG 1115+080    &1.72   &0.29  &4        &              &                       \\
 H 1413+117    &2.55   &      &4        &              &                       \\
MG 0414+0534   &2.64   &      &4        &$0.80\pm0.02$ &$71^\circ\pm5^\circ$   \\
 B 1422+231    &3.62   &      &4        &$0.73\pm0.13$ &$121^\circ\pm15^\circ$ \\
  0142--100    &2.72   &0.49  &2        &$0.71\pm0.05$ &$73^\circ\pm5^\circ$   \\
 B 0218+357    &(0.96) &0.68  &2        &              &                       \\
\hline
\end{tabular}
\end{center}
Notes -- The source and lens redshifts are $z_l$ and $z_s$ respectively. The number of images
is $N_{im}$.  The axis ratio $q_l$ and position angle of the major axis (P.A.) are 
given with their uncertainties if known.  The ellipticities and position angles are
from Falco et al. (1996) for MG 0414+0534, Impey et al. (1996) for B 1422+231, and
Falco et al. (1996) for 0142--100.  
\end{table}

Singular isothermal ellipsoids in an external shear generically produce 2, 3, 4, and 6 image 
systems depending on the amplitudes and relative orientations of the shears.  
The only systems we see are the two-image and four-image morphologies 
(with another image trapped and demagnified in the singular core).  
The three-image
cusp morphology is produced by an exposed cusp caustic and consists
of three images offset to one side of the lens galaxy (see Wallington \&
Narayan 1993 or Kassiola \& Kovner 1993 for diagrams of the image 
geometries), and the six-image
morphology is a four-image system near a cusp with two of the images
associated with the cusp split into double images.  The three-image
cusp morphology is associated with large shears or ellipticities
(or large core radii for non-singular lenses, see Kassiola \& Kovner 1993), 
and the six-image morphologies are associated with nearly perpendicular
shears and ellipticities.  

A circular SIS lens has a total multiple imaging cross section of $\pi b^2$,
and one simplification afforded by the model is that all
cross sections consist of the circular cross section multiplied by a 
dimensionless function of the shear and ellipticity.  To compute the
magnification bias we need the cross section as a function of magnification,
not just the total cross section.  Let 
$\sigma_n(>M,f,b,\epsilon,\gamma,\Delta\theta)$ be the integral cross 
section for a lens with parameters $b$, $\epsilon$, $\gamma$, and 
$\Delta \theta$ to produce a total magnification greater than $M$
subject to some limit on the flux ratios of the images $f$. 
Because of the scaling of the SIS lens, we can write
$\sigma_n(>M,f,b,\epsilon,\gamma,\Delta\theta)=
 \pi b^2 \hat{\sigma}_n(>M,f,\epsilon,\gamma,\Delta\theta)$
where $\hat{\sigma}_n$ does not depend on the mass scale of the lens $b$.
In most cases we will average over the relative orientations of the
two shear terms, and the angle averaged cross section is specified by 
$\sigma_n(>M,f,b,\epsilon,\gamma)=\pi b^2 \hat{\sigma}_n(>M,f,\epsilon,\gamma)$

We assume a selection function that detects all images with flux ratios 
between the brightest and faintest images smaller than $f$ with a 
circular critical radius in the range $\theta_{min} < 2 b < \theta_{max}$. 
For the distribution of galaxies, we assume a Schechter (1976) function 
exponent of $\alpha=-1$ and a ``Faber-Jackson'' exponent of $\gamma_{FJ}=4$ to 
describe the number counts of galaxies and the relation between luminosity 
and the velocity dispersion of the isothermal sphere,
\begin{equation}
  { dn \over dL}=
    { n_* \over L_* } \left[ { L \over L_* } \right]^\alpha \exp(-L/L_*)
  \qquad\hbox{and}\qquad 
  { L \over L_* } = \left[ { \sigma \over \sigma_*} \right]^{\gamma_{FJ}},
\end{equation}
where $n_* = (0.61\pm0.21)h^3 10^{-2}$ Mpc$^{-3}$ is the local comoving 
density of E and S0 galaxies (Loveday et al. 1992, Marzke et al. 1994), and
$\sigma_* = (220\pm20) \kms$ is the (dark-matter) velocity dispersion of an
$L_*$ galaxy (Kochanek 1993, 1994, Breimer \& Sanders 1993, Franx 1993).
The optical depth to lensing for SIS lenses in flat 
cosmologies is $\tau = \tau_* (D_{OS}/r_H)^3/30$ (Turner 1990).
The optical depth scale is
$\tau_* = 16\pi^3 n_* r_H^3 (\sigma_*/c)^4 \Gamma[1+\alpha+4/\gamma_{FJ}]
  = 0.024\pm0.012$,
where the uncertainties are dominated by $n_*$ and $\sigma_*$.
We perform all calculations in an $\Omega_0=1$ cosmological model.

If the probability distribution of the ellipticity parameters is
$dP/d\gamma d\epsilon$ and we assume that the orientations of the
external shear and the major axis of the galaxy are uncorrelated,
then the probability that a source of flux $F$ is lensed to produce $n$ 
images is
\begin{equation}
P_n(F) = \tau_* D_{OS}^3 
          \int_0^1 dx x^2(1-x)^2 \int d\epsilon d\gamma { dP \over d\gamma d\epsilon}
          \hat{\sigma}_n (\gamma,\epsilon) B_n(F,\gamma,r) 
          \left( e^{-u_{min}} - e^{-u_{max}} \right)
\end{equation}
where $x=D_{OL}/D_{OS}$, 
$u_{min} = \Delta\theta_{min}^2/\Delta\theta_*^2 (1-x)^2$ and 
$u_{max} = \Delta\theta_{max}^2/\Delta\theta_*^2 (1-x)^2$ specify the 
detectable range of separations where the characteristic image separation is
$\Delta\theta_* = 8\pi(\sigma_*/c)^2=2\parcs8 (\sigma_*/220\kms)^2$ 
and we assumed that $\alpha=-1$ and $\gamma_{FJ}=4$.  The 
magnification bias function is
\begin{equation}
   B_n(F,\gamma,r) = \left[ { dN \over d \ln F } \right]^{-1}
            \int_0^\infty d M { d P_n \over d M }(f,\epsilon,\gamma) 
             { d N \over d \ln F } \left( { F \over M } \right)
\end{equation}
where $dN/d\ln F $ is the logarithmic number counts distribution of the sources
at fixed redshift.  The magnification probability distributions $dP_n/dM$ are
computed by standard numerical methods (Kochanek \& Blandford 1987, Kochanek \& Apostolakis
1988, Wallington \& Narayan 1993). 
We use the quasar number counts model detailed in 
Kochanek (1996a), and the Dunlop \& Peacock (1990) pure luminosity evolution 
number counts model for flat spectrum radio sources (see Kochanek (1996b) for
a discussion of uncertainties in the radio luminosity function and its effects
on gravitational lensing).

\section{Elliptical Galaxies}

The simplest model for the origin of asymmetries in gravitational lenses assumes
that it is due to the ellipticity of the primary lens galaxy.  We treat two such
models, using either a singular isothermal ellipsoid or a singular
isothermal sphere in an external shear field.  We first discuss the expected
analytic scalings.  Next we discuss the ellipticities needed to produce the
observed distribution of image morphologies and to fit the individual lenses.
We compare the models to the observed lens galaxy where data are available.
Finally, we find the best fit parameters of a simple galaxy ellipticity distribution
for jointly fitting the number and properties of the observed lenses.

For the isothermal ellipsoid the asymptotic integral cross section 
for four-image systems is $P(>M) = 4\pi b^2\eta^2/(1-\epsilon^2)^{1/2} M^2$, and 
to lowest order in $\epsilon$ the total four-image cross section is 
$\sigma_4 = \pi b^2 \eta^2 \epsilon^2/6$ (see Kormann et al. 1994b).
The minimum magnification of a four-image system, estimated by 
$P(>M_{min}) =\sigma_4$ is $M_{min} = 24^{1/2}/\epsilon$.  For a source
at the origin behind the lens, the images form a cross on the major
and minor axes of axis ratio $q_c= 1-\epsilon/3$.   
To lowest order in the ellipticity $\epsilon = 3 (1-q_c)$, so
$\sigma_4= 3 \pi (1-q_c)^2 b^2\eta^2/2$ and $M_{min}= (8/3)^{1/2}/(1-q_c)$
for model eccentricities chosen to fit the observed ellipticity of a cruciform lens. 
The external shear model consists of an SIS lens in a quadrupole shear potential 
(an $F_{OS}$ shear term in eqn. (1)).
The asymptotic integral cross-section for four-image systems is 
$P_4(>M) = 4\pi b^2/M^2$, the total four-image
cross section is $\sigma_4 = 3\pi b^2\gamma^2/2$,and the minimum
magnification is $M_{min} = 8^{1/2}/\gamma$.  For the external shear model
to have the same four-image cross section as the ellipsoid model requires
$\gamma=\epsilon/3$,  and when $\gamma=\epsilon/3$ the four-image magnification probability
distributions of the two models are identical to lowest order in $\gamma$ and $\epsilon$. 
In a symmetric cruciform lens, the axis
ratio of the images is $q_c=(1-\gamma)/(1+\gamma)$ so $\gamma=(1-q_c)/2$. For  
the external shear model to produce the same ellipticity cruciform
image requires $\gamma=\epsilon/6$.  Fixed to the same axis ratio lens,
the external shear model has one-fourth the four-image  cross section of the ellipsoid
and twice the minimum magnification.   

Figure 3 shows the expected number of lenses in the JVAS radio survey and
the optical quasar sample for the ellipsoid model and for the external shear
model.  When we compare models as a function of $\epsilon=3\gamma$, the
results are identical in the low ellipticity limit, and then slowly
diverge for large ellipticities.  The numbers of four-image lenses exceed
the numbers of two-image lenses at moderately high ellipticities.  If 
$\epsilon \gtorder 0.73$ or $\gamma>1/3$, two of the cusps extend into the
three-image region allowing the production of the three-image cusp geometry.
Very flattened ellipsoids are dominated by the cusp image geometry (Kassiola \& Kovner 1993)
and produce diverging numbers of lenses.  The
external shear models are eventually dominated by the three-image cusp
systems, but not until $\gamma \gtorder 0.4$ (off the right edge of Figure 3). 
Thus, there is a limit to the fraction of lenses that can have the four-image
geometry of approximately 60-70\%, and the ellipticity must be very finely
tuned to reach this limit.  Low ellipticity lenses are dominated by the two-image 
geometry and high ellipticity lenses are dominated by the three-image cusp geometry.

The compact-source part of the JVAS radio sample has equal numbers of two- and 
four-image lenses,  requiring a typical ellipticity parameter of $\epsilon\sim0.7$ 
or a typical shear of $\gamma \sim 0.30$.  The small number of lenses in the
sample (two of each morphology) means that the mean ellipticity is not well
determined (see King et al. 1996, Kochanek 1996b).  In the quasar sample the greater
magnification bias increases the relative numbers of four-image systems 
compared to the radio sample.  The observed ratio of 2 four-image and 3 
two-image lenses is produced for $\epsilon\simeq 0.55$ or $\gamma\simeq 0.20$.

\begin{table}
\begin{center}
\begin{tabular}{rccrrcrr}
\multicolumn{8}{c}{Table 2: Elliptical Galaxy Models} \\
\hline
\multicolumn{1}{c}{Lens} &$N_{dof}$ &\multicolumn{3}{c}{Ellipsoid} &\multicolumn{3}{c}{External Shear} \\
    &  &$\epsilon$  &\multicolumn{1}{c}{$\theta_\epsilon$}  &\multicolumn{1}{c}{$\chi^2$} 
       &$\gamma_e$  &\multicolumn{1}{c}{$\theta_\gamma$}    &\multicolumn{1}{c}{$\chi^2$}    \\
PG 1115+080   &6 &$0.605\pm0.017$   &$ 66.6^\circ$  &150.7  &$0.126\pm0.004$  &$ 65.3^\circ$  & 82.1 \\
 H 1413+117   &4 &$0.580\pm0.014$   &$ 21.6^\circ$  &148.9  &$0.110\pm0.003$  &$ 21.7^\circ$  &142.0 \\
MG 0414+0534  &6 &$0.438\pm0.025$   &$ 79.1^\circ$  &110.7  &$0.098\pm0.007$  &$ 78.0^\circ$  &116.5 \\
 B 1422+231   &6 &$0.764\pm0.006$   &$126.8^\circ$  &124.0  &$0.261\pm0.004$  &$125.7^\circ$  & 40.3 \\
  0142--100   &0 &$0.211\pm0.141$   &$105.0^\circ$  &  0.0  &$0.069\pm0.047$  &$106.2^\circ$  &  0.0 \\
 B 0218+357   &0 &$0.111\pm0.072$   &$ 73.7^\circ$  &  0.0  &$0.035\pm0.024$  &$ 73.1^\circ$  &  0.0 \\
\hline
\end{tabular}
\end{center}
Notes -- The angles $\theta_\epsilon$ and $\theta_\gamma$ are the P.A.s of the major axis of 
 the model.  The P.A. uncertainties in the four-image lenses of $0.3^\circ$ or less are so
 much smaller than the uncertainties in any P.A. measurement for a lens galaxy (see Table 1) 
 that we do not include them.  The errors for the two-image lenses are ($+13^\circ$, $-28^\circ$)
 for 0142--100, and ($+40^\circ$, $-5^\circ$) for B 0218+357.  
\begin{itemize}
\item Related models of PG 1115+080: Kochanek (1991) found $\gamma=0.08\pm0.01$ and 
      $\theta_\gamma=66^\circ\pm6^\circ$ based on early ground based data with no lens position.
\item Related models of H 1413+117: Kochanek (1991) found $\gamma=0.11\pm0.01$ and 
      $\theta_\gamma=22^\circ\pm3^\circ$ based on early ground based data.
\item Related models of MG 0414+0534: Kochanek (1991) found $\gamma=0.08\pm0.02$ and 
      $\theta_\gamma=80^\circ\pm7^\circ$ based on early ground based data with no lens 
      position.  Falco et al. (1996a) found $\gamma=0.12\pm0.03$ and 
      $\theta_\gamma=77.4^\circ\pm0.1^\circ$.  
\item Related models of B 1422+231:  Hogg \& Blandford (1994) found a model with 
      $\gamma\simeq 0.29$
      with $\theta_\gamma \simeq 124^\circ$, and Kormann et al. (1994b) found a model with
      $\epsilon=0.71$ and $\theta_\epsilon=124^\circ$.  Our data differs in using the 
      precise position of the lens galaxy from Impey et al. (1996).  
\end{itemize}
\end{table}

\begin{figure}
\centerline{\psfig{figure=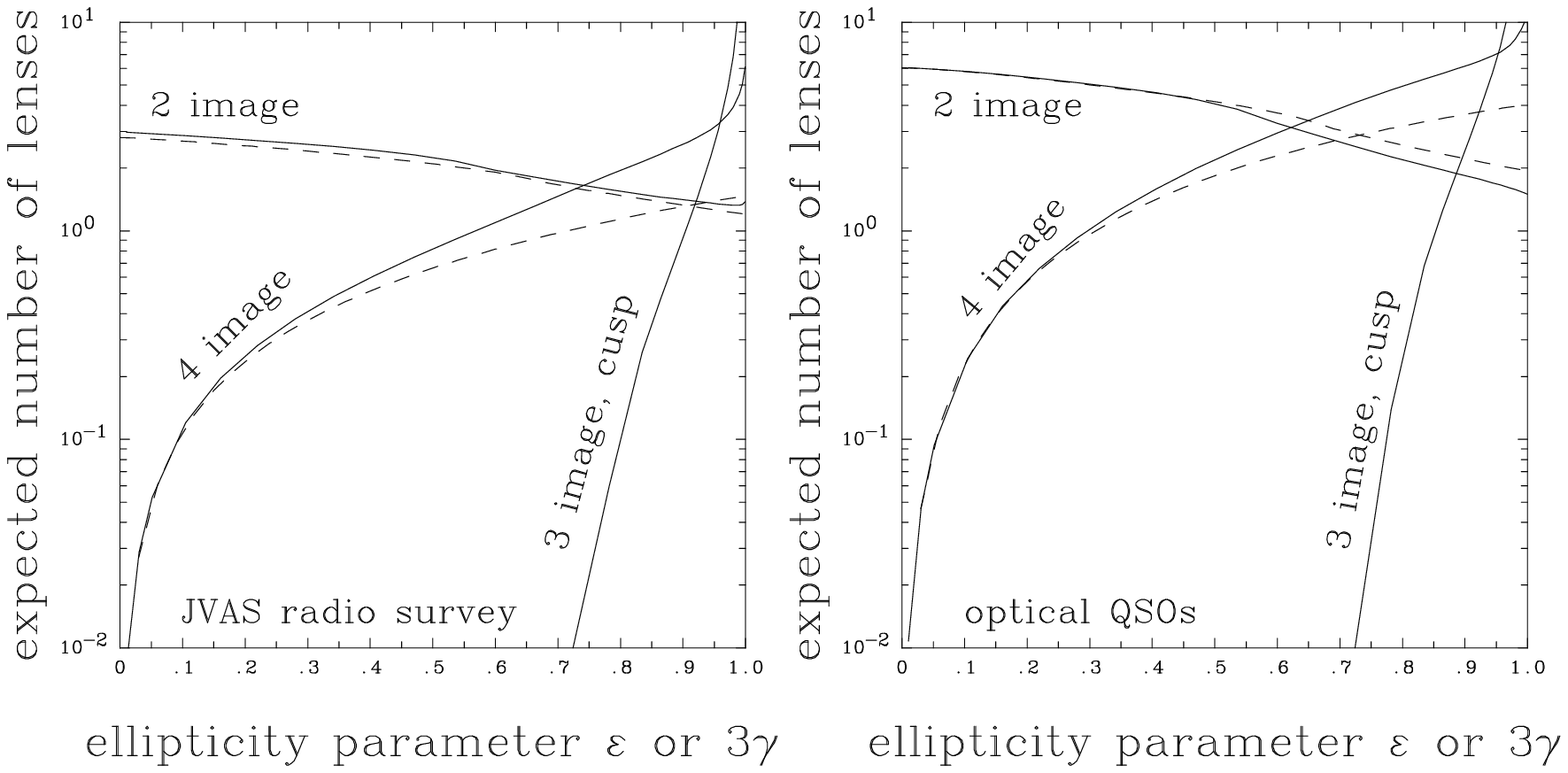,width=6in}}
\caption{The expected number of two-image, four-image, and three-image 
 cusp lenses in the JVAS radio survey (left) and the optical quasar surveys 
 (right) as a function of the ellipsoid parameter $\epsilon$ (solid) or the 
 external shear $\gamma$ (dashed).
 The horizontal scale is either $\epsilon$ or $3\gamma$ so that in the limit
 of small ellipticity the models have the same cross-sections and magnification
 probability distributions. The external shear models have a non-zero 
 three-image cusp cross section for $3\gamma > 1$.}
\end{figure}

For comparison, Table 2 summarizes the best fit models for the lenses found
in these surveys using the same two lens models.  The ellipsoid and the external 
shear models have comparable fits to the positions and flux ratios of the lenses, but
neither model provides a statistically acceptable fit to any of the four-image lenses.  The
models have only five parameters, compared to nine or eleven constraints 
for a four-image lens depending on whether there is an observed lens position, so the parameter
values of the best fitting model are well specified even if the absolute
goodness of fit is low.  Figure 4 shows the probability distributions for
the ellipticity parameter ($\epsilon$ or $\gamma$) of the models using the
same horizontal scale as in Figure 3.  
Qualitatively, the ellipticities of the 
ellipsoidal lens models are approximately equal to the ellipticities needed
to produce the observed relative numbers of two- and four-image lenses,
while the shears of the external shear models are too low.
The two-image lenses are well fit by the models due to the lack of constraints
($N_{dof}=0$!) and have broad uncertainties in their parameters.  As we
expect, the eccentricities of the two-image systems are lower than the
eccentricities of the four-image systems.
The notes to Table 2 summarize the parameters found in other identically 
parametrized models of these lenses.  The results are generally consistent,
although most parameter estimates did not include uncertainties and used older,
less accurate observational data.

For three lenses, MG 0414+0534, B 1422+231 and 0142--100, we can compare the model 
ellipticities and orientations to the images of the lens galaxy.  The models of
MG 0414+0534 (model axis ratio $0.63\pm0.02$, optical axis ratio $0.80\pm0.02$ 
(Falco et al. (1996a)) and B 1422+231 (model axis ratio $0.37\pm0.01$, optical axis 
ratio $0.80\pm0.02$ (Impey et al. 1996)) are flatter than the optical galaxy, 
while the model of 0142--100 (model axis ratio $0.80\pm0.12$, optical axis 
ratio $0.71\pm0.05$, (Falco et al. 1996b)) is rounder than the 
optical galaxy.  The B 1422+231 model is
dramatically flatter than the lens galaxy, and Hogg \& Blandford (1994) suggest 
that shear from two nearby galaxies causes the high axis ratios in single shear 
models of B 1422+231. The perturbing galaxies must be very massive to produce a total 
shear of $\gamma=0.26$.   
Changing the monopole of the lens to be more centrally concentrated
and closer to a constant mass-to-light ratio model will generally increase the
model ellipticities and exacerbate the differences (see Kochanek 1991a, Wambsganss \& 
Paczy\'nski 1994, Keeton \& Kochanek 1996b).  
If the angular structure of the lens is entirely due to an ellipsoid aligned with
the luminous galaxy, then the lens model should be aligned with the visible image
of the galaxy independent of the radial profile of the monopole.  Misalignments between the
models and the lens galaxy must be due either to external shears or misalignment
of the galaxy and its dark matter halo.  The position angles of the major axes
of the light distribution and the model differ by $8^\circ\pm5^\circ$ for
MG 0414+0534 and by $6^\circ \pm 15^\circ$ in B 1422+231.    The misalignment 
in MG 0414+0534 is significant, although smaller than the $30^\circ \pm 15^\circ$ 
misalignment observed in the X-ray halo of NGC 720 (Buote \& Canizares 1994, 1996).
For the two-image lens 0142--100 the misalignment is $32^\circ \,^{+18}_{-33}$, where
the uncertainty is dominated by the underconstrained lens model. 

\begin{figure}
\centerline{\psfig{figure=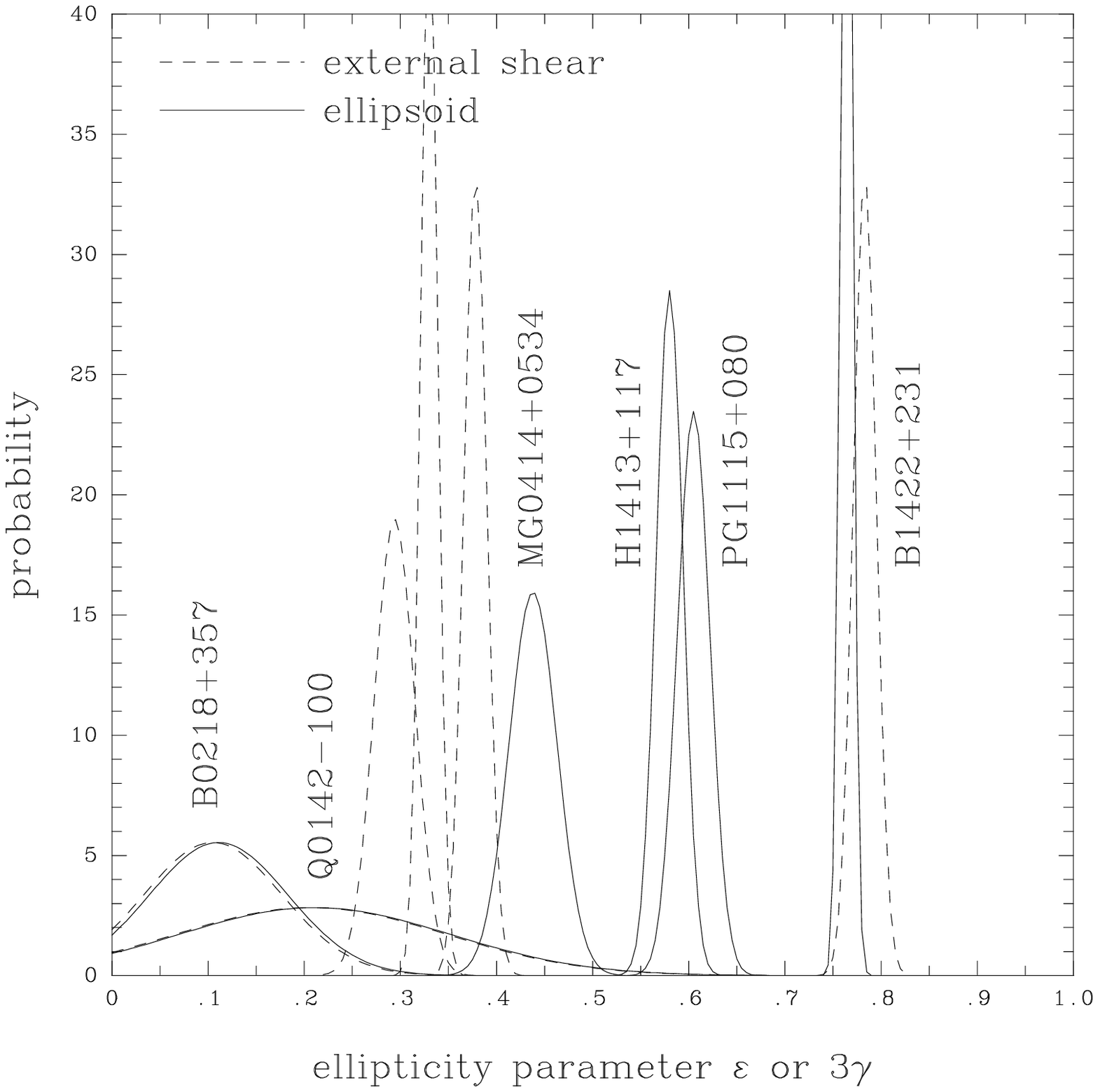,height=3.0in}}
\caption{Ellipticity $\epsilon$ (solid) or external shear $\gamma$ (dashed)
 probability distributions for the modeled lenses.
 The horizontal scale is either $\epsilon$ or $3\gamma$ so that in the limit
 of small ellipticity the models have the same cross-sections and magnification
 probability distributions.  The ordering of the lenses in shear and ellipticity
 is the same.  }
\centerline{\psfig{figure=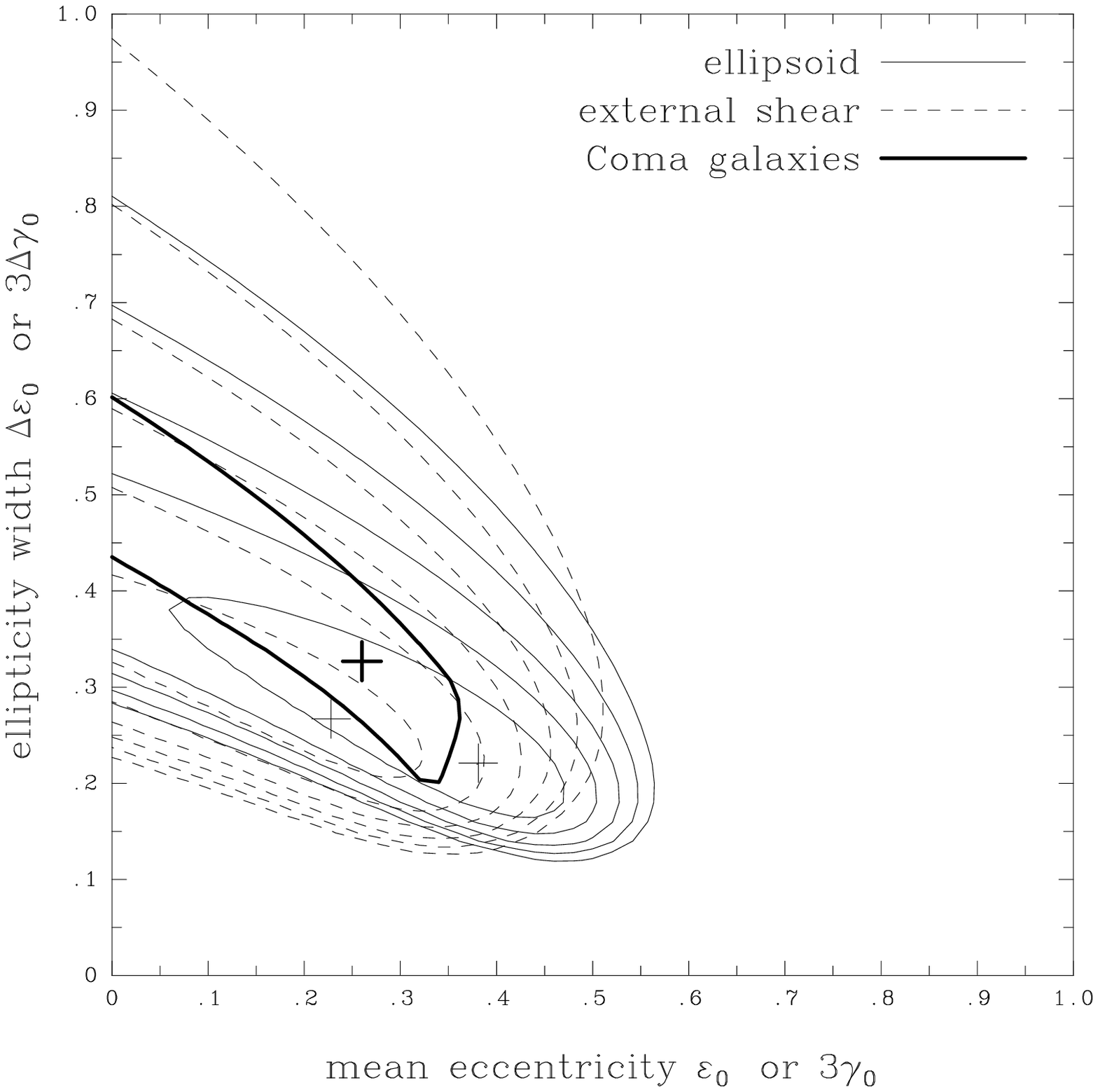,height=3.0in}
            \psfig{figure=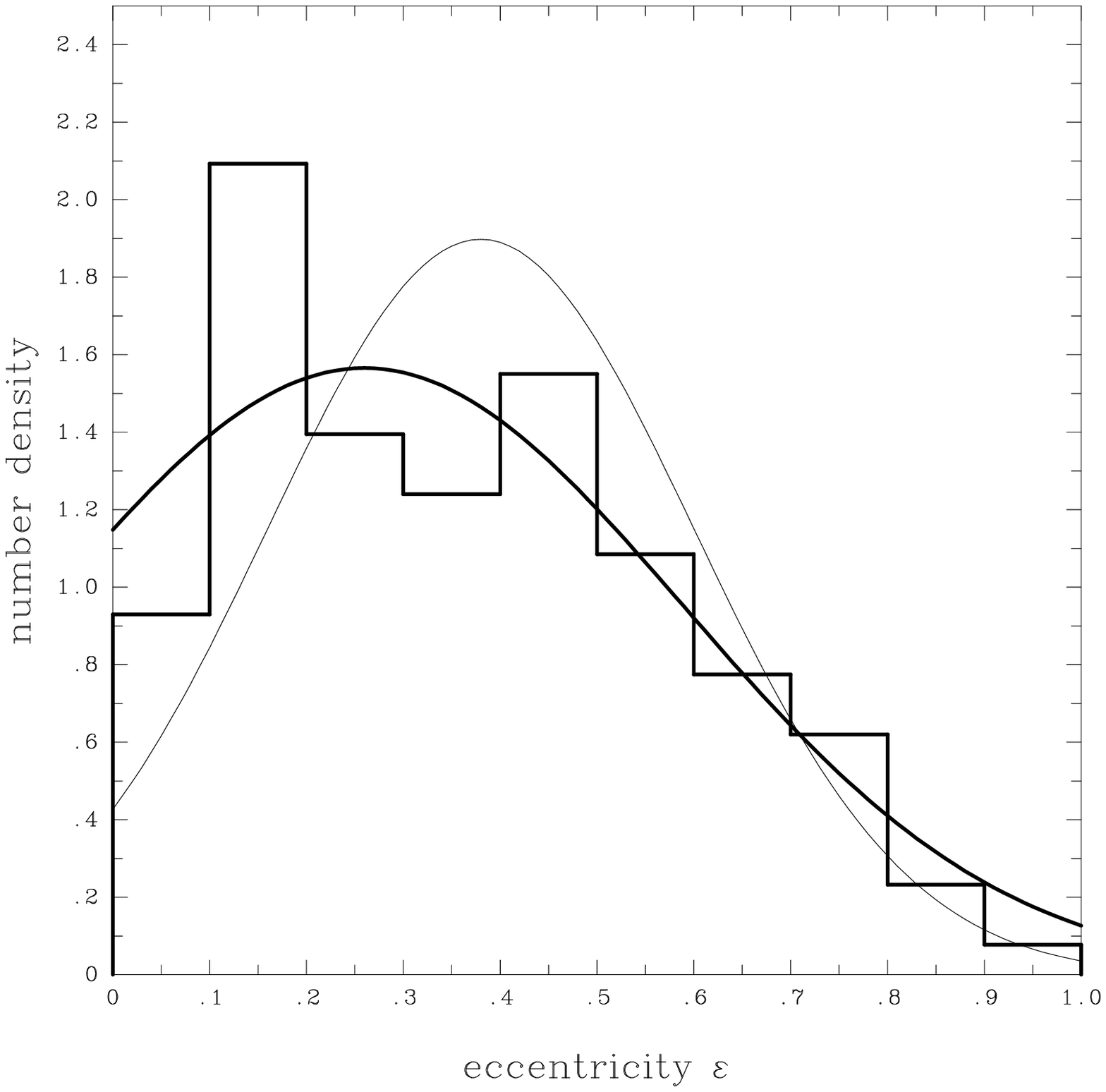,height=3.0in}}
\caption{Best fit Gaussian eccentricity (solid), external shear (dashed) 
 or Coma galaxy (heavy solid) distributions.  The left panel shows likelihood
 contours for the fits.  The crosses mark the peak likelihood models and the contours
 are spaced at 0.1 dex from the maximum.  The lowest contour is the maximum
 likelihood 90\% confidence region for two parameters. The peak likelihood
 of the external shear model is only 36\% that of the ellipsoid model. The
 heavy solid contour shows the 90\% confidence contour for the fits to the
 E+S0 galaxy sample in Coma. The right panel shows a histogram of the Coma
 sample, the best fit Gaussian eccentricity distribution to the Coma 
 sample (heavy solid), and the best fit distribution to the lens data for the
 singular isothermal ellipsoid model (light solid).  }
\end{figure}

We next assume an ellipticity distribution for the lens
galaxies and look for the maximum likelihood distribution that simultaneously
agrees with the model ellipticities and the observed numbers of lenses.
To simplify the comparison of the ellipsoid and external
shear models to the data, we assumed the same Gaussian distribution for $\epsilon$ 
used in \S2.1 to fit the observed axis ratios of galaxies. 
Figure 5 shows likelihood contours for the two parameters of fitting both
the relative numbers of two- and four-image systems as well as the observed
parameters of the lenses.  The best fit ellipsoid 
distribution is $\epsilon_0=0.38$ and $\Delta\epsilon_0=0.22$, although there 
is a broad class of acceptable solutions running to lower $\epsilon_0$ and larger 
distribution widths.  The best fit parameter values for the Coma galaxies 
from \S2.1 ($\epsilon_0=0.26$, $\Delta\epsilon=0.33$) lie well within the
90\% confidence likelihood contour at 77\% of the peak likelihood.  
Figure 5 also shows the eccentricity distribution of the Coma galaxies,
and the two best fit Gaussians. The best fit to the lens data has fewer of both 
high and low eccentricity galaxies than the fit to the observed axis ratios.    
Our use of $\eta=1$ in normalizing the lens model (see \S2.1) may mean we
are overestimating the number of high ellipticity galaxies needed to fit the 
lens data.  The best fit external shear distribution has 
$\gamma_0=0.076$ and $\Delta\gamma_0=0.089$,
again with a flat likelihood function towards lower shear and greater widths.
The best fit ellipsoid model has a significantly higher likelihood than the
best fit external shear model, with a likelihood ratio of 36\% between the
two best fitting models. The lower
relative likelihood of the external shear models is due to the mismatch between
the shear required to produce the observed numbers of four-image lenses and
the shear required to fit the lenses.

\section{Multiple Shear Models}

The ellipsoidal models of \S4 reproduce the observational data in a statistical 
sense.  A population of singular isothermal ellipsoids with the axis ratio
distribution of the early type galaxies in Coma is compatible with the lens data,
although the best fitting models are slightly more elliptical.  In our small
sample, there is no evidence for large misalignments between the models and
the observed galaxies, although the model and observed axis ratios differ.
The ellipsoidal models are more consistent with the data than the external shear
models.  There are, however, two problems.  The first problem is B 1422+231,
where the axis ratio of the lens model ($0.37\pm0.01$) is flatter than both
the observed galaxy ($0.80\pm0.02$) and the typical Coma E/S0 galaxy (6\% of the
J\o rgensen \& Franx (1994) galaxies, all S0s, are flatter than the model).
The second problem is that none of the four-image lenses is well fit by either
single shear model, and we know from Kochanek (1991a), Wambsganss \& Paczy\'nski (1994), 
and Keeton \& Kochanek (1996b) that changing the monopole structure of the lens
generally does not lead to an acceptable fit.
From the analyses in \S2, we expect all lenses to have an external 
shear perturbation in addition to the primary lens.  Here 
we consider lens models with two sources of shear, consisting
either of an ellipsoidal galaxy in an external shear field, or a LSS
shear model ($F_e$ and $F_{OL}$), and determine how the additional shear
influences the statistics and model fitting.

We first estimate the numbers of lenses of different morphologies.  Figure
6 shows the expected numbers of lenses as a function of $\gamma$ and
$\epsilon$ averaged over their relative orientations and assuming that the
shear is constant with source and lens redshift.  The average optical depth 
for four-image lenses varies with the sum in quadrature of the ellipticity 
and the shear, $\tau_4 \propto \epsilon^2 + (3\gamma)^2$, unless the primary
lens is very flattened.  In the Coma galaxy sample (J\o rgensen \& Franx 1994) 
the rms eccentricity is $\langle \epsilon^2 \rangle^{1/2} \simeq 0.43$, 
so the typical shear must be of order $\gamma\sim0.15$ before it is an
important perturbation to lens statistics.  Moreover, all the perturbative 
shear sources discussed in \S2 depend strongly on the
source and lens redshifts so the typical shears are considerably 
lower than the peak shears.  Figure 7 shows the expected number of four 
image lenses as a function of primary lens eccentricity and the maximum
rms LSS shear $\gamma_m$ (see \S2.3) for a source at $D_{OS}/r_H=1$ 
($z_s\simeq 3$).  We assumed that the two components of the effective 
shear were Gaussian random variables for fixed lens and source redshift 
with an rms effective shear of 
$\gamma_e^2 = 16 \gamma_m^2 x^2(1-x)^2 (D_{OS}/r_H)^3$.  The LSS shear 
does not significantly increase the production of four-image lenses
unless the primary lens is nearly circular ($q_2 \gtorder 0.9$).
The ratio of the four-image to the total optical depth in the Gaussian LSS 
shear model is 
$N_4/N_T \simeq \gamma_m^2 (D_{OS}/r_H)^3/14 + \langle \epsilon^2\rangle/6$,
so $\gamma_m \simeq 0.65 (D_{OS}/r_H)^{-3/2}$ is needed to make the external 
shear as important as the intrinsic ellipticity. Unless the estimates in \S2 
for the external shear ($\gamma_m \simeq 0.05(D_{OS}/r_H)^{-3/2}$)
are wrong by an order of magnitude, the lens cross sections and probabilities 
are dominated by the intrinsic properties of the primary lens galaxy.  
Similar calculations show that the other sources of external shear 
perturbations discussed in \S2.2-2.4 are also too weak to significantly 
change the expected numbers of lenses of different morphologies.

\begin{figure}
\centerline{\psfig{figure=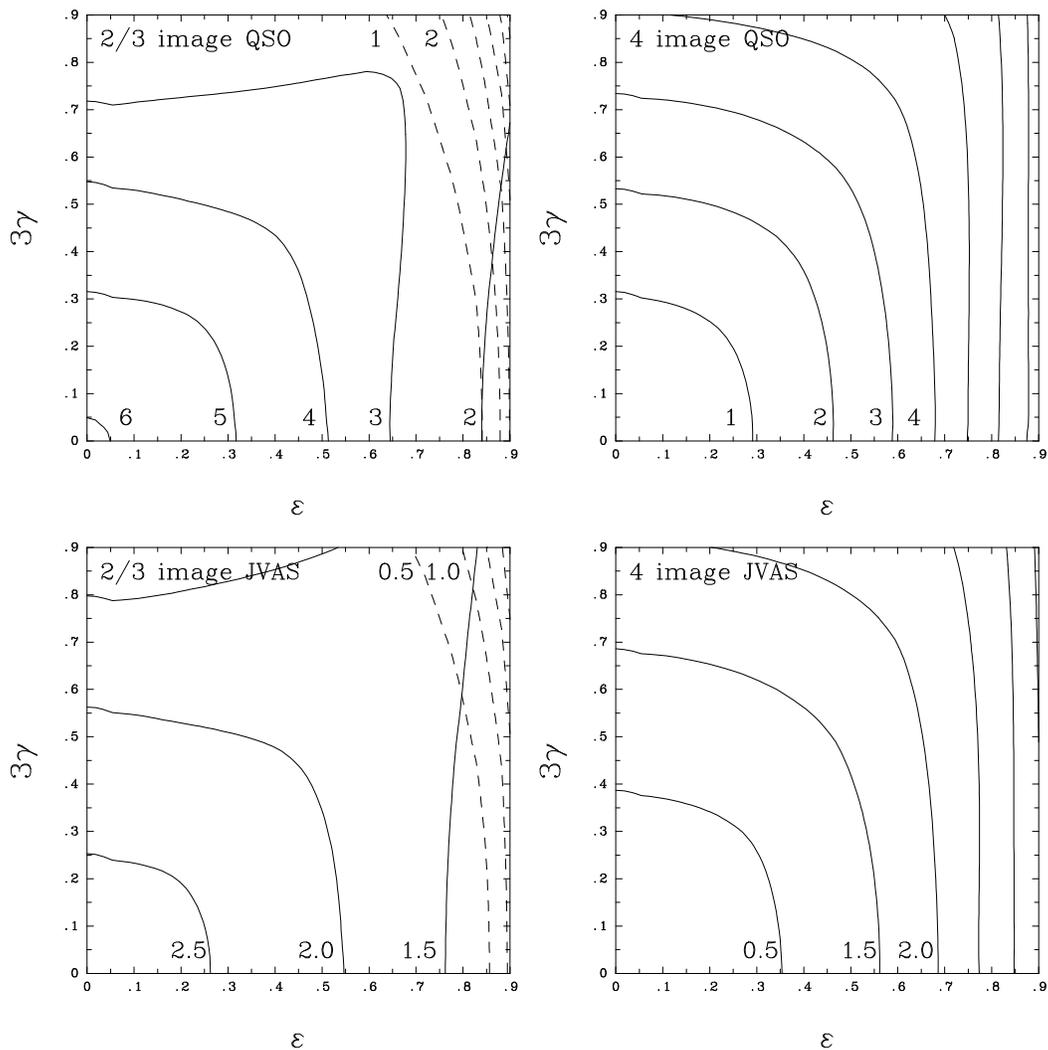,height=5.0in}}
\vspace{0.25in}
\caption{ Expected numbers of lenses in the quasar (top) and JVAS (bottom) lens
  samples as a function of the axis ratio of the isothermal ellipsoid $r$ and
  the external shear field $\gamma$.  In the 2/3 image figures, the numbers
  of two image systems are shown by the solid contours, and the number of three
  image systems by the dashed contours. The numbers label
  the nearest contour. }
\end{figure}

\begin{figure}
\centerline{\psfig{figure=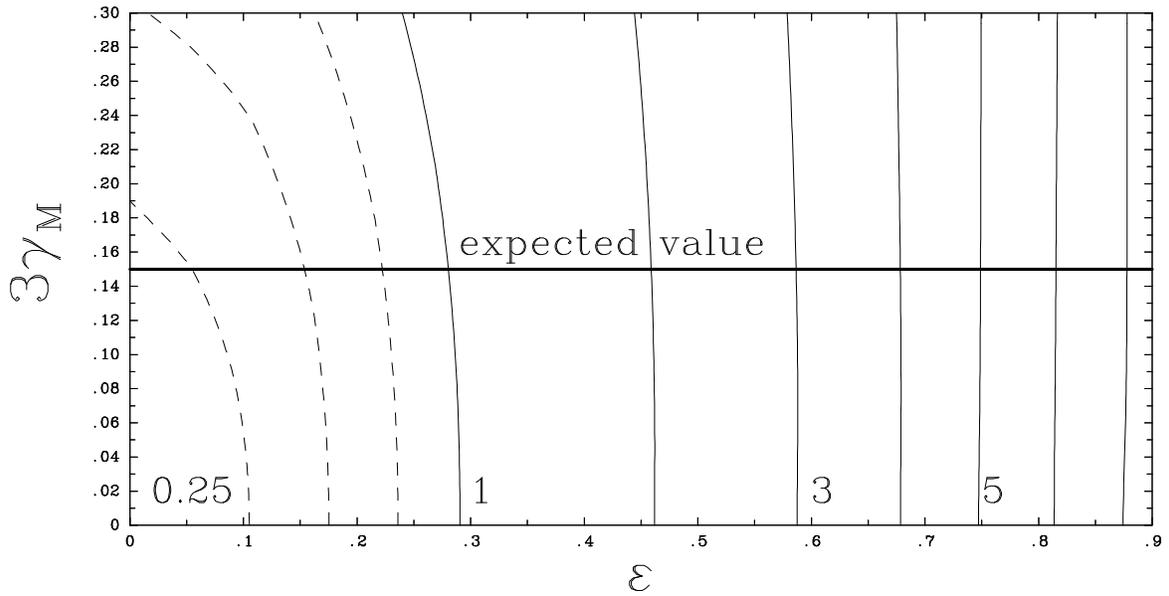,height=3.0in}}
\vspace{0.25in}
\caption{ Expected number of four-image quasar lenses as a function of 
  the maximum rms LSS shear at $D_{OS}/r_H=1$, $\gamma_m$.  The horizontal
  line marks the expected value from \S2.3.  The solid contours are spaced in steps
  of one lens, and the dashed contours in steps of 0.25 lenses.  The numbers label
  the nearest contour.
  }
\end{figure}

\begin{table}[t]
\begin{center}
\begin{tabular}{rcrccrrccr}
\multicolumn{10}{c}{Table 3: Two Shear Models} \\
\hline
  &
  &\multicolumn{4}{c}{Ellipsoid + External Shear} 
  &\multicolumn{4}{c}{SIS + LSS Shear} \\
\multicolumn{1}{c}{Lens} &\multicolumn{1}{c}{$N_{dof}$}     
    &\multicolumn{1}{c}{$\chi^2_{min}$}  &\multicolumn{1}{c}{$\epsilon$} 
    &\multicolumn{1}{c}{$\gamma$}        &\multicolumn{1}{c}{$|\Delta \theta|$} 
    &\multicolumn{1}{c}{$\chi^2_{min}$}  &\multicolumn{1}{c}{$\gamma_e$} 
    &\multicolumn{1}{c}{$\gamma_{OL}$}   &\multicolumn{1}{c}{$|\Delta \theta|$} \\
\hline
PG 1115+080    & 4  & $2.0$  &$0.22$ &$0.09$ &$36^\circ$ &$ 2.2$  &$0.12$ &$0.04$ &$40^\circ$ \\
 H 1413+117    & 2  & $0.1$  &$0.40$ &$0.19$ &$80^\circ$ &$ 0.5$  &$0.13$ &$0.12$ &$89^\circ$ \\
MG 0414+0534   & 4  &$12.1$  &$0.38$ &$0.17$ &$86^\circ$ &$48.9$  &$0.09$ &$0.06$ &$58^\circ$ \\
 B 1422+231    & 4  &$33.7$  &$0.22$ &$0.20$ &$ 3^\circ$ &$33.3$  &$0.23$ &$0.05$ &$ 4^\circ$ \\
\hline
\end{tabular}
\begin{itemize}
\item Related Models of B 1422+231:  Kormann et al. (1994b) found models with 
  parameters 
 $\epsilon=0.47$, $\gamma=0.16$ and $|\delta\theta|=4^\circ$ (model 3a) and
 $\epsilon=0.51$, $\gamma=0.10$ and $|\delta\theta|=11^\circ$ (model 3b).  
 Model 3a lies along the continuation of the $\chi^2$ ridge towards 
 higher ellipticities, while model 3b is well below the ridge.  
 Our data differs in using the 
 precise position of the lens galaxy from Impey et al. (1996).  
\end{itemize}
\end{center}
\end{table}

When we model the lenses with two shear terms we are adding two extra parameters
to the models and expect a reduction in the resulting $\chi^2$. We consider
two models, a singular isothermal ellipsoid in an external shear (in eqn. (1) 
with $F_{LS}=F_{OL}=0$), and a SIS with LSS shear (in eqn. (1) with $F_{LS}=0$,
and $\epsilon=0$).  Kochanek (1991a), Wambsganss \& Paczy\'nski (1994), and 
Keeton \& Kochanek (1996b) find that adding two extra parameters to the 
radial (monopole) structure of the lens model usually
does not dramatically improve the goodness of fit over the simple SIS + external
shear model.  Only when the lens has extended radial structure, as in the radio 
rings (see Kochanek 1995), do we expect the models to depend
strongly on the radial structure (Kochanek 1991a).
When we add two parameters to the model in the form of an additional shear term,
however, the $\chi^2$ of the model fits to our lens sample improve dramatically 
(see Table 3).  The $\chi^2$ of the PG 1115+080 and H 1413+117 models are 
statistically acceptable, the $\chi^2$ of the ellipsoid + external shear model
of MG 0414+0534 is marginally acceptable, and the $\chi^2$ of the B 1422+231 models 
and the LSS model of MG 0414+0534 are at least greatly improved.  
The two image lenses (0142--100 and B 0218+357) 
models are underconstrained (negative degrees of freedom) and were not included.  
{\it Independent of the origins of the extra shear, it is
a more fundamental variable in models of point-image lenses than variations
in the monopole structure.}  
We expect lens models to be more sensitive to multiple shear axes than statistics
for two reasons.  First, the corrections to the lens model from the extra
shear are (to lowest order) linear in the shear rather than quadratic.  
Second, models with two shear axes are qualitatively different from single
shear models, so no single shear model can generically mimic the lensing
properties of a two shear model.  Although two shear axes introduce some
qualitative changes in the statistical properties (e.g. allowing a 6 image lens),
they are related to rare and (so far) unobserved events.
We also fit the lenses with the external shear forced to be parallel or
perpendicular to the ellipsoid to test whether the improvement in the fits
could be caused by changes in the balance between the internal and external
quadrupole moments rather than misalignment.  The minimum $\chi^2$s for these
models were $83.4$, $142.5$, $108.1$, and $35.2$ for PG 1115+080, H 1413+117,
MG 0414+0534 and B 1422+231 respectively.  Except for B 1422+231, where the 
shears must be aligned to within a few degrees even in the unconstrained models,
the $\chi^2$ are much closer to the $\chi^2$ of the single ellipticity models 
of Table 2 than the unconstrained models in Table 3.
 
Most of the two shear models do not lead to unique solutions, so we list
only the model parameters at the minimum of the $\chi^2$ in Table 3.  Figure 8 
shows contours of the $\chi^2$ in the space of $\epsilon$ and $\gamma$ after 
optimizing all other parameters.  
The $\chi^2$ contours in the $\epsilon$--$\gamma$ plane generically have the
``U'' shape seen in the PG 1115+080 and H 1413+117 contours.  One side of the U, 
the ``additive branch,'' has $\epsilon + 3\gamma$ nearly constant.
Along the additive branch the angle between the major axes of the shear and
the ellipsoid is slowly varying and has no specific value ($\sim 30^\circ$
in PG 1115+080, $\sim 40^\circ$ in H 1413+117, and $\sim 2^\circ$ in B 1422+231).
Along the other two sides of the U, the ``cancellation branches,'' the shear 
and the ellipticity increase in magnitude but become perpendicular.    
The $\chi^2$ varies along the $U$, and different lenses have minima at
different points. H 1413+117 has its minimum $\chi^2$
on one of the cancellation branches with a tail extending toward the
additive branch.  MG 0414+0534 has acceptable solutions only on one of
the cancellation branches, and B 1422+231 has acceptable solutions only on
the additive branch.  The $\chi^2$ contours of the LSS models (not shown) are
qualitatively different, with the value of $\gamma_e$
nearly fixed as $\gamma_{OL}$ varies.
If we add the observed major axis of the lens as a constraint for the models
of MG0414+0534, then the best fit model has $\chi^2=66.1$ for $N_{dof}=5$,
significantly worse than for models in which the angle was free to vary.
In the unconstrained solution the PA of the major axis is  $2^\circ$, 
compared to the measured PA of $71^\circ \pm 5^\circ$.  With the constraint, the
best solution shifts to the other cancellation branch, with a galaxy PA 
of $75^\circ$, axis ratio of $0.44$ ($\epsilon=0.68$), and an external 
shear of $\gamma=0.10$.  On this branch of the solutions, the galaxy 
PA is almost exactly fit, at the price of significantly worse fits to
the positions and flux ratios.

Figure 8 also shows the integral probability distributions for either the
LSS effective shear (assuming Gaussian statistics) or the correlated shear
for each of the four lenses.  The magnitudes of the shear perturbations required 
to fit the lenses are significantly larger than the expected external shears
in all cases.  We know from the statistical models that the probability of
finding a four-image lens is not significantly enhanced by the external shear
sources discussed in \S2 unless the primary lens is nearly circular or the
estimates of the shear are gross underestimates.  While we could explain the 
high values of the secondary shears in one of these lenses using external 
shear perturbations (the best candidate being B~1422+231, Hogg \& Blandford 1994), 
it is extraordinarily unlikely that all could be explained by external shear 
perturbations.  {\it Unless \S2.2--\S2.4 grossly underestimate the typical 
external shear perturbations, the secondary shear must also be dominated 
by the primary lens galaxy rather than external perturbations.}  We defer a
discussion of the implications to the conclusions.

\begin{figure}
\centerline{\psfig{figure=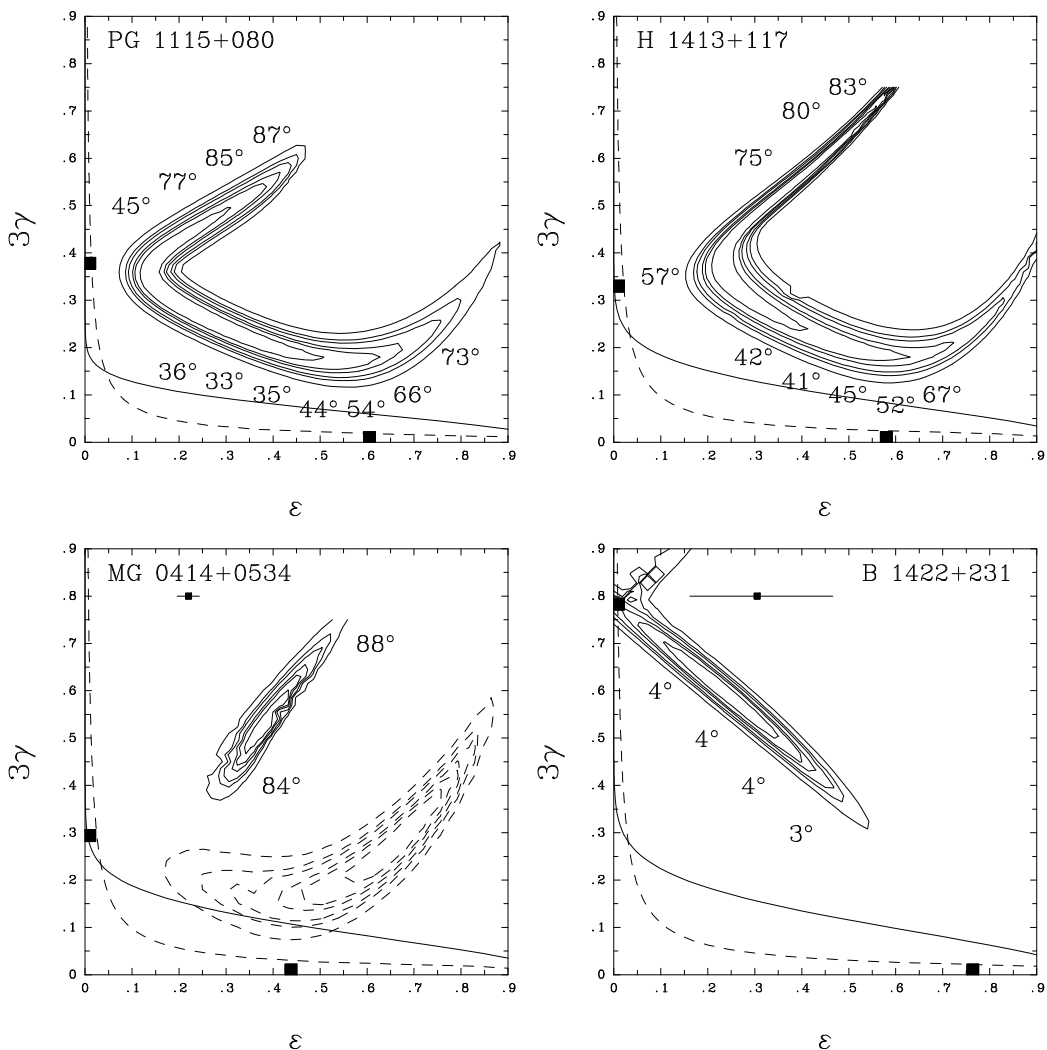,height=5.0in}}
\vspace{0.25in}
\caption{ Contours of $\Delta \chi^2$ in the plane of $\epsilon$ and $\gamma$ 
  for the four-image lenses.  Contours are drawn at $\Delta\chi^2=2.30$, 
  $4.61$, $6.17$, $9.21$, $11.8$, and $18.4$, the 1--$\sigma$, 90\%, 
  2--$\sigma$, 99\%, 3--$\sigma$, and 99.99\% confidence levels for two 
  parameters.  The models of PG 1115+080, MG 0414+0534, and B 1422+231 have 
  $N_{dof}=4$, while the model for H 1413+117 has $N_{dof}=2$ because the lens 
  position remains unknown.  The numbers give the absolute value of the angle 
  between the major axis of the ellipsoid and the shear along the minimum of 
  the $\chi^2$ function.  The heavy solid points on the axes mark the solutions
  from \S4, and the points with error bars in the MG 0414+0534 and B 1422+231  
  panels show the measured eccentricities.  The dashed contours in the 
  MG 0414+0534 panel show the $\Delta\chi^2$ contours for models with the PA 
  of the ellipsoid constrained to fit the observations.  The two curves on the 
  lower, left side of each panel
  are the integral probability of a shear exceeding $\gamma$
  for the Gaussian LSS shear model (solid) and correlated power-law shear
  model (dashed) given the redshifts of the lens.  The horizontal ($\epsilon$)
  axis becomes the integral probability.  The models assume a rms LSS shear
  of $\langle\gamma_e^2\rangle^{1/2}=0.20 x(1-x)(D_{OS}/r_H)^{3/2} $ or the 
  correlated shear model of equation (9). }
\end{figure}

If we use a single-shear model for a lens with two independent shear axes, 
one of which is aligned with the luminous galaxy, then the major axis of the
model will not be aligned with the luminous galaxy.  To estimate misalignment
angles we used the position angle of the cruciform quad lens formed by
placing the source directly behind the ellipsoid + external shear lens
model.  The misalignment angle of the images $\Delta \theta_I$ relative to 
the ellipsoid computed to first order in the eccentricity is
\begin{equation}
   \tan 2\Delta \theta_I = { 3 \gamma \sin 2 \Delta \theta \over \epsilon + 3\gamma\cos 2\Delta \theta}
\end{equation}
where $\Delta\theta$ is the angle between the ellipsoid and the external 
shear.  Figure 9 shows the mean misalignment angle and its standard deviation 
for the average lensed quasar including all variations in cross section and 
magnification bias with the angle between the shear and the ellipsoid for 
primary lens galaxies with axis ratios of $q_2 = 0.9$, $0.7$, and $0.5$.  
For $\gamma \gg \epsilon$, the distribution is uniform in the misalignment 
angle with a mean of $\langle \Delta \theta_I \rangle = 45^\circ \pm 26^\circ$.
For $\gamma \ll \epsilon$ the lens probability is independent of $\gamma$ and 
the mean misalignment approaches
$(55^\circ \pm 26^\circ)\gamma/\epsilon$ ($5.5^\circ$ for $\gamma=0.01$ and 
$\epsilon=0.1$).  The four-image lensing probability is enhanced if the shear 
is aligned with the ellipsoid, while the two-image lensing probability is 
enhanced if the shear is perpendicular, so two-image systems will show larger 
average misalignments than four-image systems for a fixed axis ratio and shear.
Unless the lens galaxy is very circular or the shear is larger than 10\%, four 
image systems should rarely show misalignments exceeding $\sim 20^\circ$.  
Interestingly, the model of the two-image lens 0142--100 shows the largest 
misalignment, although the large uncertainties make the models technically 
consistent with no misalignment.

\begin{figure}
\centerline{\psfig{figure=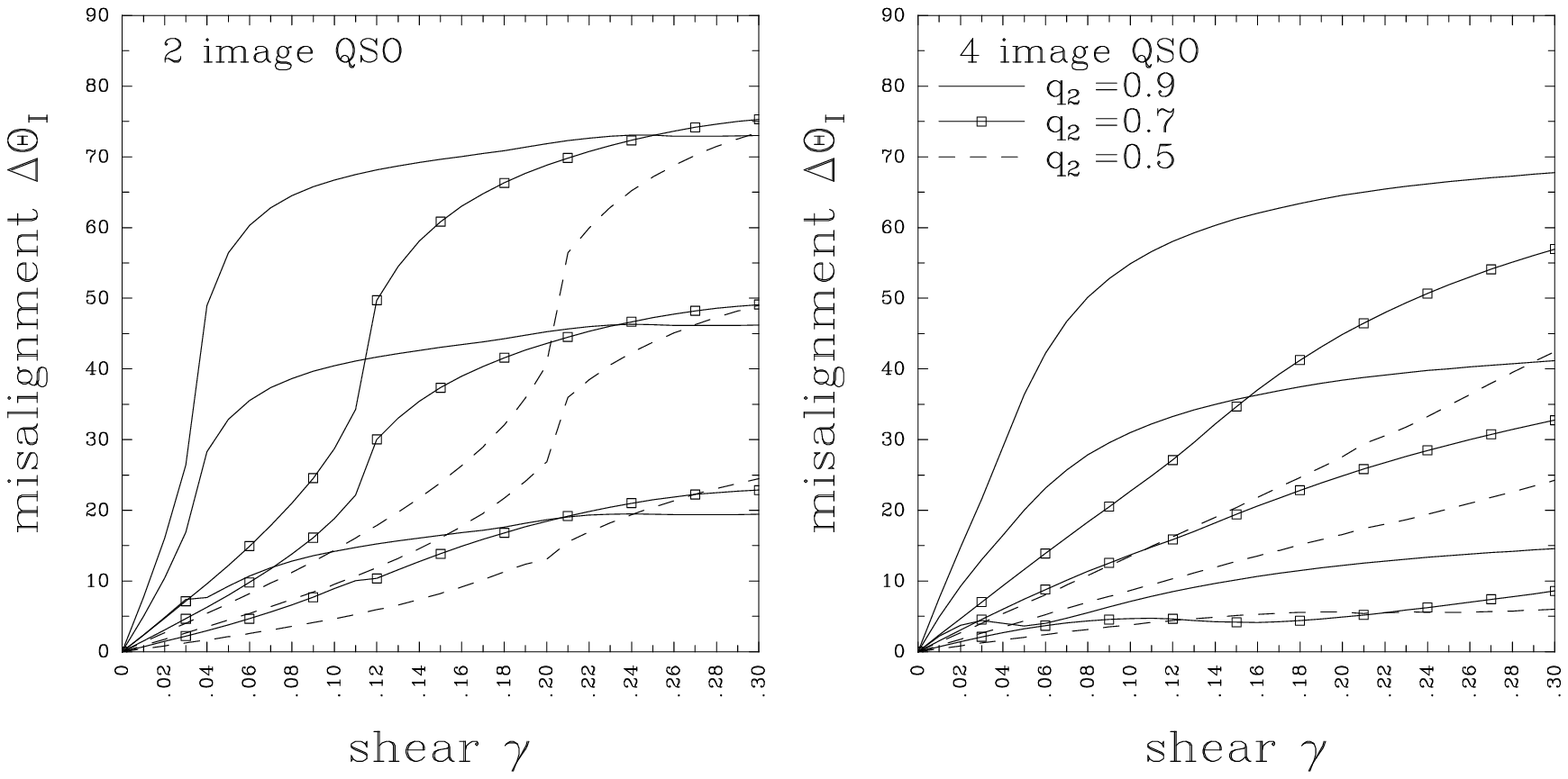,height=3.0in}}
\vspace{0.25in}
\caption{ Mean misalignment angle for two-image (left) and four-image (right)
  quasar lenses.  The solid lines, solid lines with points, and dashed lines are for primary 
  lenses with axis ratios of $q_2=0.9$, $0.7$, and $0.5$ respectively.  There are three lines
  for each axis ratio.  The central line is the mean, and the upper (lower) line is the mean
  plus (minus) one standard deviation.  A uniform random distribution in the misalignment angle
  has a mean of $45^\circ \pm 26^\circ$, and the uniform limit is reached for the high shear
  models.  External shear perturbations are expected to be smaller than 5\% for most lenses.
   }
\end{figure}

\section{Conclusions}

The most important source of ellipticity in gravitational lenses is the 
primary lens galaxy.  Early-type E and S0 galaxies produce most 
gravitational lenses, and the observed distribution of projected axis ratios is 
statistically consistent with the distribution required to produce the observed 
numbers of gravitational lenses and their ellipticities.  We assumed the presence 
of dark matter by using singular isothermal ellipsoids for the mass distribution,
because constant mass-to-light ratio models are known to be inconsistent with
the properties of gravitational lenses (Maoz \& Rix 1993, Kochanek 1995, 1996a).  
On average, the models that simultaneously produce the observed numbers of
two-image and four-image lenses and are consistent with the ellipticities
needed to model the individual lenses are somewhat
more elliptical than the light distributions of a sample of
E and S0 galaxies in the Coma cluster (J\o rgensen \& Franx 1994). 
The best fit lens models have fewer galaxies with axis ratios above $0.75$ 
and below $0.4$ than the best fit Coma model, and the normalization of our 
models may overestimate the required number of high ellipticity galaxies.  
The ellipticity of a lens model is a function of the monopole structure
(Kochanek 1991), and more centrally concentrated, constant mass-to-light ratio 
models would require higher ellipticities to fit the same data.  

In the three cases where we can compare the axis ratio of the model to the
axis ratio of the lens galaxy, one model is rounder (0142--100), one model
is somewhat flatter (MG 0414+0534), and one model is dramatically flatter
(B~1422+231, as noted earlier by Hogg \& Blandford (1994) and Kormann et al. 
(1994b)).  In B~1422+231 (Patnaik et al. 1992) the axis ratio of
our model is $0.37\pm0.01$ while the axis ratio of the galaxy is
$0.80\pm0.02$ (Impey et al. 1996), even though the major axes of the model and
the galaxy are aligned to $6^\circ \pm 15^\circ$.  A similar discrepancy
is seen in HST 14176+5226 (Ratnatunga et al. 1995, model axis ratio
$0.40$, galaxy axis ratio $0.68$).  The major axes of the 0142--100 and 
B~1422+231 models are aligned with the observed lenses to within the
model and photometric uncertainties, while the model is misaligned 
relative to the galaxy in MG~0414+0535 by $8^\circ\pm 5^\circ$.

Our simple single shear lens models do not fit the four-image
lenses very well, with typical $\chi^2/N_{dof} \simeq 20$ for both the
singular isothermal ellipsoid and SIS plus external shear models.  Studies by
Kochanek (1991a), Wambsganss \& Paczy\'nski (1994), and Keeton \& Kochanek (1996b)
show that changing the radial mass distribution of the lens causes little
improvement in the fits.  Adding an additional external shear that is 
either aligned or perpendicular to the ellipsoid is of little help in
improving the fits.  However, when we add an independent external shear that 
is not forced to be aligned with the ellipsoid, we suddenly find acceptable 
fits to most of the lenses, with $\chi^2/N_{dof} \ltorder 3$.  {\it Whatever its 
origin, an independent source of shear appears to be a more fundamental variable 
than changes in the radial mass distribution.}  An independent shear can be 
produced either by misalignments between the luminous galaxy and the dark 
matter halo, or by external shear perturbations.  

What do we expect for early-type galaxies with dark matter?  Numerical simulations 
generically find dark matter halos that are both more elliptical and more
triaxial than luminous galaxies (e.g. Warren et al. 1992, Dubinski 1992, 1994).     
Because gravitational lensing depends on the projected mass density of the
galaxy, it is far more sensitive to the properties of the halo than
most other probes of the angular structure of galaxies.  The differences
in how the mass distribution is weighted probably explain why dynamical
models of ellipticals find a mass to light ratio of $(10\pm2)h$ (e.g. van der 
Marel 1991), while lens models require $(22\pm5)h$ (Kochanek 1995, 1996).
The triaxiality of
galaxies and halos can have two important, qualitative effects on lens models.  
First, if the halos are more triaxial than the light, as expected from simulations, 
then the projected ellipticity distribution of the halos will show a larger 
deficit of low ellipticity systems than is observed in the luminous galaxies.  
In fact, the most significant difference between the estimated ellipticity distributions 
for the J\o rgensen \& Franx (1994) galaxies and the lens models is that the best 
fit lens models have smaller numbers of low ellipticity galaxies.  The lens data
requires an ellipticity distribution with fewer round galaxies to produce the
observed number of four-image lenses.  Secondly, if the triaxialities of the 
luminous and the halo matter differ, then the projected halo mass distribution 
and the projected luminosity
distribution can be misaligned, producing two different shear axes.  
In the Franx et al. (1991) study of kinematic misalignments between the 
major axes the projected rotation axes of elliptical galaxies, 26\%
of galaxies show misalignments exceeding $30^\circ$.  Comparable misalignments
can appear between the two projected mass distributions, particularly for the 
lower ellipticity lens galaxies.  Differing triaxialities
for the luminous matter and the dark matter should also appear as misalignments
between optical and X-ray isophotes, such as the $30^\circ\pm15^\circ$ 
misalignment seen in NGC 720 (Buote \& Canizares 1994, 1996).

The angular structure of the primary lens is not, however, the only source of
asymmetries contributing to gravitational lensing.  Objects correlated with the 
primary lens galaxy or near the line of sight add additional external shear
perturbations.  We estimate that the most important sources of external perturbations
are galaxies within a correlation length of the primary lens.  The shear
contribution from clusters is comparable only if every galaxy is in a cluster.
The lowest mass groups and clusters contribute most of the lensing cross section
both for multiple imaging (see Kochanek (1995b), Wambsganss et al. (1995)) and
external shear perturbations.  In the the observed lens sample, the only convincing 
lens associated with a cluster, 0957+561 (Young et al. 1980), is in a small, diffuse cluster. 
The PG 1115+080 lens galaxy appears to be part of a small group of galaxies 
(Young et al. 1981), and the external shear perturbation required to produce
a good lens model is consistent with shear from the group (Schechter et al. 1996).
The next most important source of external perturbations are other
galaxies and clusters along the line of sight (Kochanek \& Apostolakis 1988,
Jaroszy\'nski 1991).  Large external shear perturbations are associated with
discreet objects, usually galaxies, and the perturbing galaxies must be close to the
primary lens and will usually have similar or brighter fluxes.
The universe is optically thick to shear perturbations of a few percent
produced by large scale structure  (Kaiser 1992, Seljak 1994, Bar-Kana 1996),
and the rms shear fluctuations predicted using non-linear power spectra 
(Bar-Kana 1996) are quantitatively similar to the shear predicted by adding up 
the effects of discrete non-linear objects.  The probability distribution
of the high shear perturbations is a power law, and not the Gaussian 
distribution assumed in the LSS models.
For a source at a redshift of $z_s=3$, the 
typical (rms) shear is approximately 3\%, and about 5\% of lenses will
have shear perturbations exceeding 10\%.  
 
Such small external shear perturbations have negligible effects on the statistics
of gravitational lenses, because ellipsoidal mass distributions are 
more efficient than external shears in producing four-image gravitational lenses.  
The typical shear perturbation would have to be an order of magnitude larger
than predicted to significantly modify the cross sections and lensing probabilities,
and  such large shears would be trivially detected in ellipticity correlation function
experiments (Mould et al. 1994, Fahlman et al. 1994). Nonetheless,
external shear perturbations can be important for models of individual lenses.  
In particular, lenses such as B~1422+231 that require
bizarrely flattened galaxies to fit the data, probably must have strong
external shear perturbations.  Blandford \& Hogg (1994) have shown that there
are bright galaxies near B~1422+231 that can produce the necessary shear.
Even so, B~1422+231 is a joint effort between the primary lens galaxy and 
the external shear, because the major axis of the galaxy also has the
orientation needed to fit data (Impey et al. 1996).  However, since all
of our two-shear models required shear perturbations larger than expected
for external perturbations, we believe that misalignments due to dark matter
halos must be the primary origin of the secondary shear.

\acknowledgments Acknowledgements:
The authors would like to thank R. Bar-Kana, E. Falco, J. Lehar, E. Ostriker, and P. Schechter 
for discussions about ellipticity in gravitational lenses and for reading parts of the manuscript.
E. Falco kindly supplied data on several of the lenses in advance of publication.  C.S.K.
is supported by NSF grant AST-9401722.

\end{document}